\documentclass[12pt]{amsart}
\usepackage{graphicx}
\usepackage[utf8]{inputenc}
\usepackage[english]{babel}
\usepackage{enumerate,enumitem}
\usepackage{amsthm} 
\usepackage{amsmath}
\usepackage{amssymb}
\usepackage{amsmath}
\usepackage{stackrel}
\usepackage{xcolor}
\usepackage[margin=1in]{geometry}
\usepackage[normalem]{ulem}
\usepackage{hyperref}
\usepackage{tikz}
\usetikzlibrary{positioning}

\usepackage{tikz-cd}
\usepackage{adjustbox}

\usepackage{nicematrix}

\theoremstyle{plain}
\newtheorem{thm}{Theorem}[section]
\newtheorem{theorem}[thm]{Theorem}
\newtheorem{lemma}[thm]{Lemma}
\newtheorem{corollary}[thm]{Corollary}

\newtheorem{proposition}[thm]{Proposition}

\theoremstyle{definition}
\newtheorem{definition}[thm]{Definition}

\newtheorem{example}[thm]{Example}
\newtheorem{remark}[thm]{Remark}
\newtheorem{notation}[thm]{Notation}

\newtheorem{setup}[thm]{Setup}
\newtheorem{hypothesis}[thm]{Hypothesis}

\newcommand{\zerothOrderRxnsSpace}{zero-order reactions }

\def\S{\mathcal S}
\def\C{\mathcal C}
\def\Re{\mathcal R}

\title[Linear relations at steady state] {Linear relations at steady state in reaction networks}
\author[Dickenstein]{Alicia Dickenstein}
\address{Universidad de Buenos Aires and CONICET}
\author[P\'{e}rez Mill\'an]{Mercedes P\'{e}rez Mill\'{a}n}
\address{Universidad de Buenos Aires and CONICET}
\author[Shiu]{Anne Shiu}
\address{Texas A\&M University}
\date{\today}

\begin{document}

\begin{abstract}
We present a general framework for finding (affine) linear relations satisfied at steady state in chemical reaction networks. 
This framework provides a useful tool for reducing the number of variables that appear in the analysis of steady states. 
We illustrate our results on two well-studied models from systems biology; both models  pertain to the regulation of apoptosis (programmed cell death) by families of caspases.

\vskip 0.1in


\vskip 0.1in


\end{abstract}
\maketitle

\section{Introduction}
Our work focuses on the following questions concerning reaction networks taken with mass-action kinetics: What (affine) linear relations appear at steady state? Can we predict them from the network structure? How can we find them?  
Our interest in these questions comes from the fact that linear relations can replace steady-state equations, which are typically nonlinear; 
and such a replacement usually makes it easier to compute and analyze steady states.  Indeed, in this work, we illustrate our results and the way they help us understand the geometry of steady states in relevant models from systems biology.

Before describing our main results, let us explain our motivation for analyzing steady states.  Given a dynamical system arising from systems biology, we generally would like to fully understand all possible transient dynamics.  However, this task is typically too challenging, due to the nonlinearity of the system, the many parameters and variables, and so on.  Therefore, an important first step is to analyze the steady states as these are the limit points of convergent trajectories and their instability is associated with oscillations.

In order to describe our results, we provide some basic background~\cite{feinberg-book}.  A {\em chemical reaction network} consists of finitely many {\em reactions}, like $X_1+X_2 \to X_3$.  One network, which will serve to illustrate our main results, is shown later in Figure~\ref{fig:eissing-model}.  A chemical reaction network generates a system of ordinary differential equations (ODEs) by way of {\em mass-action kinetics}; an instance of such ODEs is shown later in~\eqref{eq:Eissing-ODE}. 
Solutions to these ODEs track the concentrations $x_i$  of each chemical species $X_i$; and {\em steady states} are vectors 
of concentrations at which the right-hand sides of the ODEs vanish. As all derivatives vanish at a steady state, these are the constant solutions to the ODE system.

A major research direction concerns understanding the number and stability of steady states arising from chemical reaction networks (we point the reader to a selection of works~\cite{banaji-feliu,  ME_semiopen, DPST, feliu-helmer, JoshiShiu, kothamachu-unlimited, obatake2024newton}).  A related research direction pertains to the question of what happens to the number of steady states when a network is modified in some way (for instance, by adding reactions)~\cite{Banaji, species, add-flow, ME_entrapped, ME_semiopen}.  More broadly, there has been a focus in recent years on analyzing what properties of networks (for instance, multistationarity or periodic orbits) are preserved when components of the networks (for instance, flow reactions or intermediate complexes) are added or removed~\cite{banaji-inheritance, Simp, deficiency-operation}.
This article is closely related to this circle of ideas, in that our focus is on understanding what happens to the geometry of steady states (specifically, simple sets of equations that define the set of steady states) when certain types of reactions are added or removed.

To provide more detail, 
the types of reactions we add or remove are as follows.
An {\em inflow reaction} is a reaction of the form $0 \to X_i$, while an {\em outflow reaction} has the form $X_i \to 0$.  
A {\em flow reaction} is a reaction that is an inflow reaction or an outflow reaction.
We need to consider, more generally, {\em zero-order reactions} (like $0 \to 2X_1$) and {\em first-order reactions} (like $X_1 \to X_2 + X_3$) which we now recall:

\begin{definition}\label{def:flow}
 A {\bf zero-order reaction} (respectively, {\bf first-order reaction}) is a reaction in which the reactant complex is the zero complex (respectively, is monomolecular). 
 \end{definition}
 
Our main result, Theorem~\ref{thm:main}, gives general answers concerning the occurrence of linear relations at steady state in chemical reaction networks, based on subnetworks obtained by deleting zero-order and first-order reactions.  In particular, we show that such linear relations are generated by linear conservation relations of subnetworks.  

The linear relations we obtain are, in fact, examples of the ``steady-state invariants'' Gunawardena 
introduced for the purpose of non-parametric model selection~\cite{Gunawardena-distributivity}, with follow-up work by many researchers~\cite{case-study, complex-linear-invariant, MacLean-Wnt}.
In this context, our linear relations provide variables that, at steady state, must lie on a linear space, regardless of the parameter values.  Thus, if there are several competing models, steady-state data obtained experimentally can be used to reject or support a candidate model, without knowing the parameter values (such parameter values are often difficult to estimate in models arising in systems biology).

\smallskip

As an application of our results, we analyze the steady states in two models of apoptosis (programmed cell death).  
Apoptosis is triggered by the activation of certain specialized enzymes, called {\em caspases}, that digest essential cellular constituents and lead to degradation of genomic DNA~\cite{albeck2008,cohen1997caspases}. 
We focus on two well-studied models of caspase activation.  The first model, which has $8$ species, was introduced and validated by Eissing {\em et al.}\ in 2004~\cite{eissing2004bistability}, with follow-up investigations by Eissing and co-authors~\cite{eissing-chaves-all,Waldherr}
and others~\cite{analyse-eissing}. The second model we consider is more detailed than the Eissing {\em et al.}\ model; it has $58$ species and was proposed by Albeck {\em et al.}\ in 2008~\cite{albeck2008}.

\begin{table}[ht]
\begin{tabular}{lcccc}
\hline
Model  & {\small \# vars.} & \# {\small linear cons. laws} & \# {\small steady-state relns.} & {\small new \# vars.} \\
\hline
Eissing {\em et al.} & 8   & 0        & 4       & 4     \\
Albeck {\em et al.}  & 58  & 18       & 15      & 25    \\
\hline      
\end{tabular}
\caption{Summary of application of our results to two models of caspase activation in apoptosis.  Displayed for each model are the number of (species) variables, the number of (linearly independent) linear conservation laws, the number of (linearly independent) steady-state relations arising from Theorem~\ref{thm:main}, and the resulting decreased number of variables.  For details, see Example~\ref{ex:eissing-linear} and Section~\ref{sec:Albeck}.}
\label{tab:caspase-summary}
\end{table}

As summarized in Table~\ref{tab:caspase-summary}, our results allow us to greatly reduce the number of variables appearing in the analysis of steady states.  To give some detail, we show that while the first model does not have any linear conservation laws, the steady states live in a four-dimensional linear variety, which reduces the number of variables from $8$ to $4$. As for the second model, in addition to $18$ (linearly independent) linear conservation laws, there are $15$ additional linear relations satisfied at steady state, which reduces the number of variables from $58$ to $25$. 

To give more ideas behind our main result and its applications, let us delve briefly into the Eissing {\em et al.} model.  
The reaction network underlying this model is shown in Figure~\ref{fig:eissing-model}.  The $8$ species variables are given in Table~\ref{tab:eissing-species}, and the biological meaning behind the reactions is shown in Table~\ref{tab:eissing-reactions}.  The resulting ODEs (arising from mass-action kinetics) are as follows, where the terms arising from zero-order or first-order reactions (i.e., monomials of degree-0 or degree-1 in the variables $x_i$) are marked in red: 
{\footnotesize
 \begin{equation}\label{eq:Eissing-ODE}
	\begin{split}
\frac{dx_1}{dt} ~&=~ - k_2 x_1 x_4 \textcolor{red}{-  k_9 x_1 + k_{-9}} 
\\
\frac{dx_2}{dt} ~&=~ 
- k_{11} x_2 x_7 +  k_2 x_1 x_4 \textcolor{red}{-  k_5 x_2 + k_{-11} x_8} \quad \quad 
\\
\frac{dx_3}{dt} ~&=~ 
 - k_1 x_2 x_3 \textcolor{red}{-  k_{10} x_3 + k_{-10}}
 \\
\frac{dx_4}{dt} ~&=~ 
  k_1 x_2 x_3 -  k_3 x_4 x_5 \textcolor{red}{-  k_6 x_4 + k_{-3} x_6}
	\end{split}
	\begin{split}
\frac{dx_5}{dt} ~&=~  - k_3 x_4 x_5 -  k_4 x_4 x_5 \textcolor{red}{-  k_8 x_5 + k_{-3} x_6 + k_{-8}}
\\
\frac{dx_6}{dt} ~&=~   k_3 x_4 x_5 \textcolor{red}{-  k_7 x_6 - k_{-3} x_6}
\\
\frac{dx_7}{dt} ~&=~  - k_{11} x_2 x_7 \textcolor{red}{-  k_{12} x_7 + k_{-11} x_8 + k_{-12}}
\\
\frac{dx_8}{dt} ~&=~   k_{11} x_2 x_7 \textcolor{red}{-  k_{13} x_8 - k_{-11} x_8}
	\end{split}    
    \end{equation}
}   

\begin{figure}

\begin{center}
\begin{tikzpicture}[node distance=0.7cm and 0.5cm,scale=0.7]

\node (23) {Cas8*+Cas3};
\node (24) [below right=0.1cm and 0.7cm of 23] {Cas8*+Cas3*};
\node (14) [below left=0.1cm and 0.7cm of 24] {Cas8+Cas3*};
\node (45) [below=of 14] {Cas3*+XIAP};
\node (6) [right=of 45] {Cas3*-XIAP};
\node (out6) [right=of 6] {$0$};
\node (4) [below=of 45] {Cas3*};
\node (out4) [right=of 4] {$0$};
\node (27) [below=of 4] {Cas8*+CARP};
\node (8) [right=of 27] {Cas8*-CARP};
\node (out8) [right=of 8] {$0$};
\node (70) [right=0 and 1cm of out8] {CARP};
\node (out7) [right=of 70] {$0$};
\node (30) [above=of 70] {Cas3};
\node (out3) [right=of 30] {$0$};
\node (10) [above=of 30] {Cas8};
\node (out1) [right=of 10] {$0$};
\node (50) [above=of 10] {XIAP};
\node (out5) [right=of 50] {$0$};
\node (20) [above=of 50] {Cas8*};
\node (out2) [right=of 20] {$0$};

\draw[->,> = latex] (23) -- (24) node[midway, above] {$k_1$};
\draw[->,> = latex] (14) -- (24) node[midway, below] {$k_2$};
\draw[->,> = latex] (45.5) -- (6.175) node[midway, above] {$k_3$};
\draw[->,> = latex,color=red] (6.-175) -- (45.-5) node[midway, below] {$k_{-3}$};
\draw[->,> = latex,color=red] (6) -- (out6) node[midway, above] {$k_7$};
\draw[->,> = latex] (45) -- (4) node[midway, left] {$k_4$};
\draw[->,> = latex,color=red] (4) -- (out4) node[midway, above] {$k_6$};
\draw[->,> = latex] (27.5) -- (8.175) node[midway, above] {$k_{11}$};
\draw[->,> = latex,color=red] (8.-175) -- (27.-5) node[midway, below] {$k_{-11}$};
\draw[->,> = latex,color=red] (8) -- (out8) node[midway, above] {$k_{13}$};
\draw[->,> = latex,color=red] (70.5) -- (out7.168) node[midway, above] {$k_{12}$};
\draw[->,> = latex,color=red] (out7.-166) -- (70.-5) node[midway, below] {$k_{-12}$};
\draw[->,> = latex,color=red] (30.5) -- (out3.168) node[midway, above] {$k_{10}$};
\draw[->,> = latex,color=red] (out3.-166) -- (30.-5) node[midway, below] {$k_{-10}$};
\draw[->,> = latex,color=red] (10.5) -- (out1.168) node[midway, above] {$k_9$};
\draw[->,> = latex,color=red] (out1.-166) -- (10.-5) node[midway, below] {$k_{-9}$};
\draw[->,> = latex,color=red] (50.5) -- (out5.168) node[midway, above] {$k_8$};
\draw[->,> = latex,color=red] (out5.-166) -- (50.-7) node[midway, below] {$k_{-8}$};
\draw[->,> = latex,color=red] (20) -- (out2) node[midway, above] {$k_5$};
\end{tikzpicture}
\end{center}

\caption{The reactions underlying the Eissing {\em et al.} model~\cite{eissing2004bistability}, with zero-order and first-order reactions marked in red.}
    \label{fig:eissing-model}
\end{figure}

\begin{table}[ht]
\begin{tabular}{llllllll}
\hline
$x_1$                  & $x_2$ & $x_3$ & $x_4$ \\
\hline
(initiator) caspase-8 &  activated caspase-8    & (effector) caspase-3 & activated caspase-3  
\\
\hline
 $x_5$ & $x_6$ & $x_7$ & $x_8$ \\
 \hline
 XIAP (inhibitor) & (caspase-inhibitor cpx.) & CARP (inhibitor) & (caspase-inhibitor cpx.)  \\
\hline
\end{tabular}
\caption{Species in the Eissing {\em et al.}\ model.  Here, XIAP is a protein whose name is short for ``X-linked inhibitor of apoptosis'', and CARP means ``CARD containing protein'' (and CARD is ``caspase-associated recruitment domain'').} 
\label{tab:eissing-species}
\end{table}

\begin{table}[ht]
\tiny
\begin{tabular}{llll}
\hline
Rate & Reaction & Rewritten & Meaning \\
\hline
$k_1$
&
Cas8* + Cas3 $\to$ Cas8* + Cas3*
&
$X_2+X_3 \to X_2+X_4$ 
& 
Activation of caspase-3 \\
\hline
$k_2$
&
Cas8 + Cas3* $\to$ Cas8* + Cas3*
&
$X_1+X_4 \to X_2 + X_4$ 
& 
Activation of caspase-8 
\\
\hline
$k_3,k_{-3}$
&
Cas3*+XIAP $\rightleftarrows $ Cas3*--XIAP
&
$X_4+X_5 \rightleftarrows X_6$ 
& 
Inactivation of caspase-3* by XIAP
 \\
\hline
$k_4$
&
Cas3*+XIAP $\to $ Cas3*
&
$X_4+X_5 \to X_4$
& 
Cleavage of XIAP by caspase-3* (products neglected)
\\
\hline
$k_5$
&
Cas8* $\to$ 0
&
$X_2 \to 0$
& 
Degradation
 \\
\hline
$k_6$
&
Cas3* $\to$ 0
&
$X_4 \to 0$
& 
Degradation
 \\
\hline
$k_7$
&
Cas3*--XIAP $\to$ 0
&
$X_6 \to 0$ 
& 
Degradation
 \\
\hline
$k_8, k_{-8}$
&
XIAP $\rightleftarrows$ 0
&
$X_5 \rightleftarrows 0$ 
& 
Degradation and synthesis
 \\
\hline
$k_9, k_{-9}$
&
Cas8 $\rightleftarrows$ 0
&
$X_1 \rightleftarrows 0$ 
& 
Degradation and synthesis
 \\
\hline
$k_{10}, k_{-10}$
&
Cas3 $\rightleftarrows$ 0
&
$X_3 \rightleftarrows 0$ 
& 
Degradation and synthesis
 \\
\hline
$k_{11},k_{-11}$
&
Cas8*+CARP $\rightleftarrows$ Cas8*--CARP
&
$X_2+X_7 \rightleftarrows X_8$ 
& 
Sequestration of activated caspase-8
 \\
\hline
$k_{12},k_{-12}$
&
CARP $\rightleftarrows$ 0
&
$X_7 \rightleftarrows 0$ 
& 
Degradation and synthesis
 \\
\hline
$k_{13}$
&
Cas8*--CARP $\to$ 0
&
$X_8 \to 0$ 
& 
Degradation
 \\
\hline
\end{tabular}
\caption{Reactions in the Eissing {\em et al.}\ model.  The degradation and synthesis reactions are the outflow and inflow reactions, respectively. }
\label{tab:eissing-reactions}
\end{table}

 By removing all zero-order and first-order reactions, we obtain the following subnetwork:
\begin{align} \label{eq:Eissing-subnetwork-smallest}
    X_1+X_4 \xrightarrow{k_2}  
    X_2 + X_4 \xleftarrow{k_1} X_2+X_3
    \quad 
    X_4 \xleftarrow{k_4} X_4+X_5 \xrightarrow{k_3} X_6 \quad 
    X_2 + X_7 \xrightarrow{k_{11}} X_8~.  
\end{align}       
While the original model has no linear conservation laws, it turns out that the above subnetwork has the following (linearly independent) conservation laws, the first three of which correspond to the total amounts of (respectively) caspase-8, caspase-3, and CARP:
\begin{align} \label{eq:Eissing-smallest-subnetwork-cons-laws}
    x_1+x_2+x_8=T_1,
    \quad
    x_3 + x_4 + x_6 = T_2,
    \quad
    x_7 + x_8 = T_3,
    \quad 
    k_3 x_5 + (k_3 + k_4) x_6 = T_4.
\end{align}
The constants  $T_1,T_2,T_3,T_4$ are usually called {\em total amounts}. 
The fourth conservation law, unlike the first three, has coefficients depending on the rate constants (specifically, $k_3$ and $k_4$). The appearance of this ``extra'' linear conservation law is due to the existence of two {\em terminal strong linkage classes}  (strongly connected components with no edges leaving them) $\{X_4\}$ and $\{X_6\}$ in the second {\em linkage class} (connected component) of the subnetwork~\eqref{eq:Eissing-subnetwork-smallest}, as predicted by~\cite[Corollary~2]{feinberg-horn-1977} (also being used here is the fact that the subnetwork~\eqref{eq:Eissing-subnetwork-smallest} is deficiency-zero).
 
It turns out that, by applying Theorem~\ref{thm:main}, 
the conservation laws~\eqref{eq:Eissing-smallest-subnetwork-cons-laws} of the subnetwork induce the following 4 (affine) linear relations (in the $x_i$) that hold at all steady states of the original Eissing model~\eqref{eq:Eissing-ODE}: 
    \begin{align}
        \label{eq:eissing-4-linear0}
        \notag
    - k_9 x_1 - k_5 x_2 -k_{13} x_8 + k_{-9} ~&=~ 0 \\
    -k_{10} x_3 - k_6 x_4 - k_7 x_6 + k_{-10} ~&=~0\\
        \notag
               -k_{12} x_7 - k_{13} x_8 + k_{-12} ~&=~0 \\
        \notag
       -k_3 k_8 x_5 -  (k_3 k_7 + k_4 k_7 + k_4 k_{-3})x_6 + k_{-8} k_3
       ~&=~0
    \end{align}
The message here is that, even though the Eissing {\em et al.} model has no linear conservation laws (no ``total amounts'' are preserved), the four linear relations~\eqref{eq:eissing-4-linear0} hold at steady state. Thus, when analyzing the original steady-state equations -- which come from setting to $0$ the eight (nonlinear) right-hand sides of the ODEs~\eqref{eq:Eissing-ODE} -- we can replace four of these equations by the simpler linear relations~\eqref{eq:eissing-4-linear0}.
We refer the reader to 
Section~\ref{sec:Eissing-model} for further details.

\smallskip

As for the EARM model of apoptosis proposed in~\cite{albeck2008}, we show that the associated mass-action system has several interesting features. As mentioned earlier, the number of variables can be reduced from $58$ to $25$ using the affine-linear relations at steady state (see Definition~\ref{def:spaces-conservation-laws}). Moreover, we prove that, while the system is conservative and so the stoichiometric compatibility classes containing the trajectories are compact, all non-negative steady states lie on the boundary (as an instance of our Lemma~\ref{lem:positive_rel}) and are degenerate, and the variety of steady states in each class can be $2$-dimensional. Also, we explain the decrease in the rank of the associated Jacobian matrices as arising from a linear relation among the gradients {(see Proposition~\ref{prop:syzygy}).

\smallskip

Going beyond models of caspase activation, we expect our results and general framework for analyzing linear steady-state relations to apply to many ODE models arising in systems biology.  To explain why, let us first provide some context.  Many networks arising in biology (e.g., in signaling pathways) take the form of a subnetwork of ``core'' reactions -- this subnetwork often has many linear conservation laws representing total amounts -- 
together with inflows and outflows (representing synthesis and degradation, respectively).  
When these networks are modeled with mass-action kinetics, as is commonly done, our results apply.  

One key message is that, even after adding flow reactions, 
which typically destroys some of the linear conservation laws, the destroyed laws transform into affine linear steady state relations (see Corollary~\ref{coro:dimensionsum} for details).  Thus, the linear conservation laws generate the same number of (linearly independent) linear relations and/or conservation laws: some of them ``extend'' to the full network, while the others transform into affine linear steady state relations.
Accordingly, our work fits into a body of work mentioned earlier that aims to understand the effect of adding flow reactions.

\smallskip

This work is organized as follows.  
We begin with a concise introduction to reaction networks in Section~\ref{sec:background}.   
We present our theoretical results in Section~\ref{sec:linear}, with the proofs postponed until Section~\ref{sec:proofs}. 
In Sections~\ref{sec:Eissing-model} and~\ref{sec:Albeck}, we apply our results to the Eissing {\em et al.} and the Albeck {\em et al.}\ models of apoptosis, respectively.
Finally, in Section~\ref{sec:discussion}, we further discuss the relevance of our work to systems biology.

\section{Brief background on chemical reaction networks} \label{sec:background}

In this section, we very briefly define reaction networks, the resulting mass-action kinetics ODEs, and steady states.  
These systems appear widely in applications; here we only refer the reader to~\cite{invitation,feinberg-book,toth-book,polly-gheorghe-review} for more details on such applications and the basic setting.

A {\bf chemical reaction network} $N=\{\S,\C,\Re\}$ consists of three finite sets:
(1)  a set of {\bf species} $\S:=\{X_1, X_2, \dots, X_n\}$;
(2) a set of {\bf complexes} $\C=\{y_1, y_2, \dots, y_m\}$, which are non-negative-integer combinations of the species; and 
(3) a set of {\bf reactions} $\Re \subseteq (\C \times \C)\smallsetminus\{(y,y)  \mid  y\in\C\}$.  A {\bf subnetwork} of a reaction network $N=\{\S,\C,\Re\}$ is a reaction network $N'=\{\S',\C',\Re'\}$ for which 
$\S' \subseteq \S,\C' \subseteq \C$, and $\Re' \subseteq \Re$.

For a reaction $(y_i,y_j)$, the {\bf reactant} is $y_i$ and the {\bf product} is $y_j$.  
As is standard, we represent the reaction $(y_i,y_j)$ by the arrow $y_i\to y_j$, as seen, for instance, in the reaction network for the Eissing {\em et al.}\ model shown earlier in Figure~\ref{fig:eissing-model}.  See also the subnetwork~\eqref{eq:Eissing-subnetwork-smallest}.

It is useful to represent a complex $a_1X_1+a_2X_2+\dots + a_nX_n$ by the vector in $(a_1,a_2,\dots,a_n)\in \mathbb{Z}^n_{\geq 0}$. For instance, the complex $X_2+X_3$ in 
the Eissing {\em et al.}\ model can be represented by $(0,1,1,0,0,0,0,0)$.
A complex $y_i \in \mathbb{Z}^n_{\geq 0}$ is {\bf monomolecular} if $y_{i1}+y_{i2}+ \dots + y_{in}=1$ (so, $X_2$ is monomolecular, but $X_2+X_3$ is not).

Next, we explain how a reaction network generates, via mass-action kinetics, a dynamical system.  Consider a reaction network $N$ with $r$ reactions.  We assign to each reaction, a positive {\bf rate constant}: for the $i$-th reaction $y_i \to y_i'$, we denote the rate constant by $k_{i} \in \mathbb{R}_{> 0}$.  
We collect these rate constants into a vector
$\kappa:=(k_1, k_2, \dots, k_r)$.  Next, 
we write the vector of species concentrations as $x=(x_1,x_2, \dots, x_n)$.  
Together $N$ and $\kappa$ define the following ODE system:
    \begin{equation}
    \label{eq:mass_action_ODE}
    \frac{dx}{dt} 
    ~=~
    \sum_{i=1}^r  k_{i} x^{y_i} (y_i'-y_i) ~=:~ f_{\kappa}(x)~,
    \end{equation}
where $x^{y_i} := \prod_{j=1}^n x_j^{y_{ij}}$.
For the Eissing {\em et al.}\ network, the ODEs~\eqref{eq:mass_action_ODE} appeared earlier in~\eqref{eq:Eissing-ODE}.

In general, the right-hand side of the ODEs~\eqref{eq:mass_action_ODE} is a vector of polynomials 
$f_{\kappa,i}(x) \in \mathbb{Z}[k_1,k_2,\dots, k_r][x_1,x_2,\dots, x_n]$, for $i=1,2, \dots,n$. For convenience, we often write $f_i$ in place of $f_{\kappa,i}$.  
The {\bf reaction vector} of a reaction $y_i \to y_i'$ is the difference $y_i'-y_i$. 
The right-hand side of the ODEs~\eqref{eq:mass_action_ODE} always lies in the {\bf stoichiometric subspace} 
\begin{equation} \label{eq:defS}
S:= \mathrm{span} \{y_i'-y_i  \mid~ i=1,2,\ldots, r\} ~,
\end{equation}
which is the linear subspace spanned by the reaction vectors. It follows from~\eqref{eq:mass_action_ODE} that for any vector $\lambda$ in the orthogonal subspace $S^\perp$, the equality $\sum_{i=1}^n \lambda_i f_i=0$ holds.

It follows that the solution $x(t)$ to the ODE system~\eqref{eq:mass_action_ODE} with initial condition $x(0)$ remains in the coset $x(0)+S$.  In fact, the non-negative orthant is invariant for the dynamics, so if $x(0) \in \mathbb{R}^n_{\geq 0}$, then $x(t)$ remains non-negative and thus stays in the following {\bf stoichiometric compatibility class}, for all time $t$ in any interval  containing $0$ where $x$ is defined:
\begin{align*}
    \mathcal{P} ~=~ (x(0)+S) \cap \mathbb{R}^n_{\geq 0}~. 
\end{align*}

The orthogonal complement of $S$ defines linear forms that vanish on $S$ and are therefore linear conservation laws. 
In the next section, we will extend the concept of conservation law to also allow relations with coefficients that depend on the rate constants (but not on $x_1,x_2,\dots, x_n$).
We will say more in Remark~\ref{rem:stoichiometry}.

A {\bf steady state} of a mass-action system~\eqref{eq:mass_action_ODE} with $n$ species is a concentration vector $x^*\in \mathbb{R}^n$ at which the right-hand side of the ODEs~\eqref{eq:mass_action_ODE} vanishes: $f_{\kappa}(x^*)=(f_1(x^*),f_2(x^*),\ldots,f_n(x^*))=0$. In the setting of chemical reaction networks, the focus is on non-negative steady states $x^*\in \mathbb{R}^n_{\geq 0}$.
A steady state $x^* \in {\mathbb R}^n_{\geq 0}$ is a {\bf boundary steady state} if $x^*_i=0$ for some $i$.  

When there are no linear conservation laws, a steady state $x^*$ is  {\bf degenerate} if  the Jacobian matrix $J_f$ of $f_1,f_2, \dots, f_n$ has determinant $0$ at $x^*$.  In general, a steady state $x^*$ is {\bf degenerate} when
${\rm ker}(J_f(x^*)) \cap S \not= \{0\}$.
In particular, if the rank of $J_f(x^*)$ is less than $\dim(S)$, then $x^*$ is degenerate.

\section{Linear relations at steady state} \label{sec:linear}

This section contains our mathematical results. We also illustrate our results through examples that clarify our definitions, hypotheses, and statements. The main setting for this section is detailed below in  our Setup~\ref{setup:star} and Definition~\ref{def:spaces-conservation-laws}.  Two important applications of our results appear later, in Sections~\ref{sec:Eissing-model} and~\ref{sec:Albeck}. Our proofs are postponed to Section~\ref{sec:proofs}.

\subsection{Our setting}\label{ssec:setting}

The following is the setting and notation of our results.

\begin{notation}
For a vector of unknown rate constants $\kappa:=(k_1,k_2,\dots, k_r)$, we use the following standard notation for the rings of polynomials with integer coefficients and for rational functions with rational coefficients (that is, quotients of polynomials with rational coefficients where the denominator is not the zero polynomial):
    \[
     \mathbb{Z}[\kappa]~ := ~ 
        \mathbb{Z}[k_1,k_2,\dots, k_r]
            \quad {\rm and}\quad 
    \mathbb{Q}(\kappa)~ := ~ 
        \mathbb{Q}(k_1,k_2,\dots, k_r)~.
    \]
\end{notation}

\medskip

\begin{setup}
    \label{setup:star}
    We assume that $N$ is a reaction network obtained (as a subnetwork) from a reaction network $\widetilde{N}$ by removing one or more zero-order and/or first-order reactions (as in Definition~\ref{def:flow}).
  We denote by $n$ the number of species of $\widetilde{N}$; we also view $N$ as having $n$ species, even if some species do not participate in any reaction of $N$.  Let $r$ and $r+p$ denote the number of reactions of $N$ and $\widetilde{N}$, respectively.  
   We denote the mass-action ODE systems arising from $N$ and $\widetilde N$, respectively, as follows:
    \[
    \dot x = f^N(x)
    \quad \quad {\rm and}
    \quad \quad \dot x = f^{\widetilde{N}}(x)~.    
    \]
    Here, for $i=1,2,\dots, n$, we have $f^N_i \in \mathbb{Z}[\kappa][x_1,x_2,\dots, x_n]$ and $f^{\widetilde{N}}_i \in \mathbb{Z}[\widetilde{\kappa}][x_1,x_2,\dots, x_n]$,  where the rate constants $\kappa=(k_1,k_2,\dots, k_{r})$ and $\widetilde{\kappa}=(k_1,k_2,\dots, k_{r+p})$ are viewed as unknowns. 
   For instance, if $\widetilde{N}$ is the network shown earlier in Figure~\ref{fig:eissing-model}, and $N$ is the subnetwork~\eqref{eq:Eissing-subnetwork-smallest}, then $\widetilde{\kappa}=(k_1,k_2,\dots, k_{13}, ~k_{-3},k_{-8},k_{-9},k_{-10}, k_{-11},k_{-12})$ and $\kappa=(k_1, k_2, k_3, k_4, k_{11})$.
    To alleviate the notation, we simply write $f$ instead of $f^{\widetilde{N}}$ if there is no confusion.
\end{setup}

\medskip

\begin{definition}[Steady-state relation]
\label{def:spaces-conservation-laws}
Let $N$ be a reaction network with $n$ species and $r$ reactions. Consider $\dot x = f^N(x)$ the mass-action ODE system arising from $N$.
Here, for $i=1,2,\dots, n$, we have $f^N_i \in \mathbb{Z}[\kappa][x]:= \mathbb{Z}[\kappa][x_1,x_2,\dots, x_n]$, where the rate constants $\kappa=(k_1,k_2,\dots, k_r)$ are viewed as unknowns.

Consider a  polynomial $h$ in $ \mathbb{Q}(\kappa)[x]:= \mathbb{Q}(\kappa)[x_1,x_2,\dots,x_n]$:
\[
    h~=~ \sum_{\alpha \in A} b_{\alpha} x^{\alpha}~,
\]
where $A$ is a finite subset of $\mathbb{Z}_{\geq 0}^n$, each coefficient $b_{\alpha}$ is in $\mathbb{Q}(\kappa)$ (for $\alpha \in A$), and, as usual, we use the notation $x^{\alpha}:=x_1^{\alpha_1} x_2^{\alpha_2} \dots x_n^{\alpha_n}$ for $ \alpha\in \mathbb{Z}_{\geq 0}^n$.
\begin{itemize}
        \item The polynomial $h $ is a {\bf steady-state relation} of $N$ if, for all $\kappa^*$ such that the denominators of the coefficients $b_{\alpha}$ do not vanish at $\kappa^*$, if $x^*$ satisfies $f^N(\kappa^*, x^*) =0$, then  $h(\kappa^*, x^*) =0$.
        \item The polynomial $h$ is an {\bf affine-linear steady-state relation} of $N$ if $h$ is a steady-state relation of $N$ and, additionally, $h$ is affine-linear in the unknowns $x_i$ (that is, $h$ is of the form $h= a_0 +  a_1x_1 + a_2 x_2 + \dots + a_n x_n$, for some $a_0,a_1, \dots, a_n \in \mathbb{Q}(\kappa)$).
    \end{itemize}
    We further define two vector spaces arising from $N$:
    \begin{enumerate}
         \item The {\bf space of linear conservation relations} of $N$ is: 
        \begin{align*}
            L_{\kappa}(N) ~:=~
                \left\{ 
                c \in \mathbb{Q}(\kappa)^n \mid c_1 f^N_1 + c_2 f^N_2 + \dots + c_n f^N_n = 0  
                \right\}~.
        \end{align*}
        Elements of $L_{\kappa}(N)$ are also called {\bf linear conservation laws} because it follows from the equality $\dot{x} = f^N(x)$ that, over any interval of definition of the functions $x_1,x_2, \dots, x_n$, there exists $c_0$ that is independent of $x_1,x_2, \dots, x_n$ such  that $c_1 x_1 + c_2 x_2 + \dots + c_n x_n = c_0 $.
        \item The subspace of {\bf affine-linear steady-state relations} that are obtained {\em linearly} from the polynomials $f^N_i$ is as follows:
        \[ A_{\kappa}(N) ~:=~ 
        \{ h \in \mathbb{Q}(\kappa)[x] \text{ affine-linear } \mid~
        h = d_1 f^N_1 + d_2 f^N_2 + \dots + d_n f^N_n~\text{ with }d_i \in \mathbb{Q}(\kappa)
        \}~.\]
   \end{enumerate}
\end{definition}

An example of a network $N$ with an affine-linear steady-state relation that is {\em not} in $A_{\kappa}(N)$ (i.e., not obtained linearly from the polynomials $f^N_i$)
appears at the end of this section; see Example~\ref{ex:one-target}.

\begin{remark}[Subscripts]
    \label{rem:notation-kappa}
In Definition~\ref{def:spaces-conservation-laws}, we use the subscript $\kappa$ in the notation $L_{\kappa}(N)$ 
to remind the reader that $L_{\kappa}(N)$ is a subset of $ \mathbb{Q}(\kappa)^n $.
In particular, when we have two networks $N$ and $\widetilde{N}$, we typically let $\widetilde{\kappa}$ denote the rate constants of $\widetilde{N}$ and then 
$L_{\widetilde{\kappa}}(\widetilde{N})$ is a subset of $ \mathbb{Q}(\widetilde{\kappa})^n $.
\end{remark}

\begin{notation}
    \label{notation:N-subnet-N-tilde}
When $N$ is a subnetwork of $\widetilde{N}$ (with vectors of rate constants $\kappa$ and $\widetilde{\kappa}$, respectively), we consider the following spaces, which generalize those in Definition~\ref{def:spaces-conservation-laws}:
 \begin{align*}
            L_{\kappa}(\widetilde{N}) ~:=~
                \left\{ 
                c \in \mathbb{Q}(\kappa)^n \mid c_1 f^{\widetilde{N}}_1 + c_2 f^{\widetilde{N}}_2 + \dots + c_n f^{\widetilde{N}}_n = 0  
                \right\}~, \quad {\rm and}
        \end{align*}
       \[ A_{\kappa}(\widetilde{N}) ~:=~ 
        \{ h \in \mathbb{Q}(\widetilde{\kappa})[x] \text{ affine-linear } \mid~
h = d_1 f^{\widetilde{N}}_1 + d_2 f^{\widetilde{N}}_2 + \dots + d_n f^{\widetilde{N}}_n~\text{ with }d_i \in \mathbb{Q}(\kappa)
        \}~.\]
\end{notation}

\begin{remark}[Linear conservation relations from stoichiometry]
    \label{rem:stoichiometry}
The space of linear conservation relations $L_{\kappa}(N)$ contains all linear conservation relations arising from stoichiometry, by which we mean the linear forms (with integer coefficients) vanishing on the {\em stoichiometric subspace $S$} in~\eqref{eq:defS}. 
An instance of such conservation relations are the ``total amount'' conservation laws shown earlier in~\eqref{eq:Eissing-smallest-subnetwork-cons-laws}.

Typically, linear forms vanishing on $S$ are the only  linear conservation laws of the system $\dot{x}=f(x)$.   
To clarify when $\dot{x}=f(x)$ has additional linear conservation laws, involving coefficients that depend on the rate constants~$\kappa$, we refer the reader to the work of Feinberg and Horn~\cite{feinberg-horn-1977}.     
\end{remark}

\begin{remark}[Computing $L_{\kappa}(N)$]
\label{rem:compute-linear-conservation-relations}
For any network $N$, the space of linear conservation relations, $L_{\kappa}(N)$, can be computed via standard techniques from linear algebra as follows.  
Starting from the right-hand sides of the mass-action ODEs, replace each monomial (in the variables $x_i$)  with a new variable, use software to do Gaussian elimination (over $\mathbb{Q}(\kappa)$), and then revert each new variable to its original monomial.
\end{remark}

In light of Remark~\ref{rem:compute-linear-conservation-relations}, a natural question is whether, for a given reaction network $\widetilde{N}$, the space 
$A_{\widetilde{\kappa}}(\widetilde{N})$
can be computed efficiently.  Our main result will answer this question: we can compute $A_{\widetilde{\kappa}}(\widetilde{N})$
via the space of linear conservation relations $L_\kappa(N)$ of a suitable subnetwork~$N$
(see Theorem~\ref{thm:main} and Proposition~\ref{prop:generates} in the next subsection).

Before proceeding to our main results, we illustrate our definitions and preview our results through a well-known model; see, for instance,~\cite{WangSontag}.

\begin{example}[Futile cycle]
    \label{ex:futile-cycle}
     Consider the following ``futile cycle'' (1-site phosphorylation network), which we call $\widetilde{N}$:
      \begin{align*}
        S_0+E  \stackrel[k_{-1}]{k_{1}}{\rightleftarrows} Y_1 \xrightarrow{k_2} S_1 +E, 
        \quad 
        S_1+F  \stackrel[k_{-3}]{k_{3}}{\rightleftarrows} Y_2 \xrightarrow{k_4} S_0+F. 
    \end{align*}  
We delete all reactions with reactant complex $Y_1$ or $Y_2$ (i.e., all first-order reactions), which yields the following network which we denote by $N$:
 \begin{align*}
        S_0+E \xrightarrow{k_1} Y_1, 
        \quad 
        S_1+F \xrightarrow{k_3} Y_2. 
    \end{align*}  
The mass-action ODEs of $N$ are as follows (where we use the usual notation $[S_0], \dots, [F]$ instead of $x_1, \dots, x_6$):  
\begin{align*}
        \frac{d[S_0]}{dt} ~&=~ 
	-k_1[S_0][E]
        ~=:~
        f^N_{S_0}
        \quad &
        \frac{d[S_1]}{dt} ~&=~ 
	-k_3[S_1][F]
        ~=:~
        f^N_{S_1} 
        \\
        \frac{d[Y_1]}{dt} ~&=~ 
	+k_1[S_0][E]
        ~=:~
        f^N_{Y_1} 
        \quad &
        \frac{d[Y_2]}{dt} ~&=~ 
	+k_3[S_1][F]
        ~=:~
        f^N_{Y_2} 
        \\
        \frac{d[E]}{dt} ~&=~ 
	-k_1[S_0][E]
        ~=:~
        f^N_{E} 
        \quad &
        \frac{d[F]}{dt} ~&=~ 
	-k_3[S_1][F]
        ~=:~
        f^N_{F} ~,
    \end{align*}
while those for $\widetilde{N}$ are:  

    \begin{align} \label{eq:ode-futile-cycle}
        \notag
        \frac{d[S_0]}{dt} &= 
        f^N_{S_0} + 
            k_{-1} [Y_1] + k_4 [Y_2]
        ~=:~
        f^{\widetilde{N}}_{S_0}
        \quad &
        \frac{d[S_1]}{dt} &= 
        f^N_{S_1} 
            + k_2 [Y_1] + k_{-3} [Y_2]
        ~=:~
        f^{\widetilde{N}}_{S_1}
        \\
        \frac{d[Y_1]}{dt} &= 
        f^N_{Y_1} 
        - (k_{-1}+k_2) [Y_1] 
        ~=:~
        f^{\widetilde{N}}_{Y_1}
                \quad &
        \frac{d[Y_2]}{dt} &= 
        f^N_{Y_2} 
        - (k_{-3}+k_4) [Y_2] 
        ~=:~
        f^{\widetilde{N}}_{Y_2}
        \\
        \notag
        \frac{d[E]}{dt} &= 
        f^N_{E} 
        +(k_{-1}+k_2) [Y_1] 
        ~=:~
        f^{\widetilde{N}}_{E}
        \quad &
        \frac{d[F]}{dt} &= 
        f^N_{F} 
        + (k_{-3}+k_4) [Y_2] 
        ~=:~
        f^{\widetilde{N}}_{F}~.
    \end{align}
It is straightforward to check that $N$ has the following $4 $ (linearly independent) linear conservation laws, which represent 
the total amounts of (respectively) unphosphorylated substrate~$S_0$, phosphorylated substrate~$S_1$, kinase~$E$, and phosphatase~$F$:
    \begin{align} \label{eq:conservation-law-subnetwork-of-1-site}
    f^N_{S_0} + f^N_{Y_1} ~=~ 0~, \quad
    f^N_{S_1} + f^N_{Y_2} ~=~ 0~, \quad
    f^N_{E} + f^N_{Y_1} ~=~ 0~, \quad
    f^N_{F} + f^N_{Y_2} ~=~ 0~. 
    \end{align}
In contrast, the larger network $\widetilde{N}$ has only $3$ (linearly independent) linear conservation laws, representing the total amounts of substrate $S$, kinase $E$, and phosphatase~$F$:
    \begin{align} \label{eq:conservation-law-1-site}
    f^{\widetilde{N}}_{S_0} + f^{\widetilde{N}}_{Y_1} +
        f^{\widetilde{N}}_{S_1} + f^{\widetilde{N}}_{Y_2} ~=~ 0~, \quad
    f^{\widetilde{N}}_{E} + f^{\widetilde{N}}_{Y_1} ~=~ 0~, \quad
    f^{\widetilde{N}}_{F} + f^{\widetilde{N}}_{Y_2} ~=~ 0~.
    \end{align}
The conservation laws~\eqref{eq:conservation-law-subnetwork-of-1-site}--\eqref{eq:conservation-law-1-site}
can be recast as follows:
    \begin{align} \label{eq:lin-cons-law-space-1-site}
        L_{\kappa}(N) ~& = ~ \mathrm{span}_{\mathbb{Q}(\kappa)} \left( \{ 
        e_{S_0} + e_{Y_1},~ 
        e_{S_1} + e_{Y_2},~ 
        e_{E} + e_{Y_1}  ,~ 
        e_{F} + e_{Y_2} \} \right)
\quad 
{\rm and}
\\
\notag
    L_{\widetilde{\kappa}}(\widetilde{N}) ~& = ~ \mathrm{span}_{\mathbb{Q}(\widetilde{\kappa})}\left(\{     e_{S_0} + e_{Y_1}+
        e_{S_1} + e_{Y_2}  ,~ 
    e_{E} + e_{Y_1}  ,~ 
    e_{F} + e_{Y_2} \} \right)~,
    \end{align}
where $e_E, e_F,\dots$ denote canonical basis vectors and we use the notation $\kappa=(k_1,k_3)$ and $\widetilde{\kappa}=(k_1,k_2,k_3,k_4,k_{-1},k_{-3})$.

Returning to the form of the conservation laws in~\eqref{eq:conservation-law-subnetwork-of-1-site}, we observe that the third and fourth conservation laws (those for the kinase and phosphatase) both extend from $N$ to $\widetilde{N}$, as seen in the second and third conservation laws in~\eqref{eq:conservation-law-1-site}.  But what happens to the other two conservation laws, which represent unphosphorylated and phosphorylated substrate, respectively?  
One answer is that they somehow combine to represent the total amount of substrate (in phosphorylated or unphosphorylated form), as in the first conservation law in~\eqref{eq:conservation-law-1-site}.  
But what if we were to examine the corresponding two ``non-extending conservation laws'', which simplify (via~\eqref{eq:ode-futile-cycle}--\eqref{eq:conservation-law-subnetwork-of-1-site}) as follows:
    \begin{align}
        \label{eq:new-relation-1-site}
            f^{\widetilde{N}}_{S_0} + f^{\widetilde{N}}_{Y_1} ~=~
            k_4[Y_2]-k_2[Y_1]~
            \quad {\rm and} \quad 
            f^{\widetilde{N}}_{S_1} + f^{\widetilde{N}}_{Y_2} ~=~
            -k_4[Y_2]+k_2[Y_1]~.
    \end{align}
Observe that the expression $h:=k_4[Y_2]-k_2[Y_1]$ is affine linear (in the species variables $[Y_1]$ and $[Y_2]$). Also, as seen in~\eqref{eq:new-relation-1-site}, $h$ is a sum of two of the $f^{\widetilde{N}}_i$ and thus vanishes at steady state.   We conclude that $h$ is an element of $A_{\widetilde{\kappa}}(\widetilde{N})$.

\smallskip

Several natural questions arise:  
    \begin{enumerate}
        \item Does $A_{\widetilde{\kappa}}(\widetilde{N})$ contain steady-state relations besides $h$?  More precisely, is $A_{\widetilde{\kappa}}(\widetilde{N})$ generated by $\{h\}$?  
        \item Does the procedure we followed -- taking the ``non-extending conservation laws'' of a subnetwork $N$ -- always yield affine-linear steady state relations?  
    \end{enumerate}
We answer both questions in the affirmative in Theorem~\ref{thm:main} and Proposition~\ref{prop:generates} below. \end{example}

\subsection{Statements of our results}\label{ssec:results}

In this subsection, we state our main results. All 
proofs are postponed until Section~\ref{sec:proofs}. 

\medskip

We will use the notation and Setup~\ref{setup:star} from Subsection~\ref{ssec:setting}. 

    \medskip

The following theorem and its corollaries relate the concepts introduced in Definition~\ref{def:spaces-conservation-laws} and Notation~\ref{notation:N-subnet-N-tilde}.

\begin{theorem}[Steady-state relations obtained from subnetwork's conservation laws] 
\label{thm:main}
Consider networks $N$ and $\widetilde{N}$, as in Setup~\ref{setup:star}. 
    For $i=1,2,\dots,n$, consider the difference 
    \[
    \ell_i ~:=~ f_i^{\widetilde{N}}
    - f^N_i ~\in~ \mathbb{Z}[\widetilde{\kappa}][x_1,x_2,\dots, x_n]~,\] 
    so that the ODE system for $\widetilde{N}$ is as follows, where $\ell(x):=(\ell_1(x),\ell_2(x), \dots, \ell_n(x))^{\top}$:
    \begin{align} \label{eq:lemma-full-system}
    \dot x ~&= ~ f^{\widetilde{N}}(x) ~=~f^N(x) +  \ell(x)~.
        \end{align}

Then, the following hold:
\begin{enumerate}
    \item \label{it:affine-linear}
    For all $i=1,2,\dots,n$, the expression $\ell_i $ 
    is affine linear in $x$.
    \item \label{it:phi}
    If $c=(c_1,c_2,\dots,c_n) \in \mathbb{Q}(\kappa)^n$ defines a linear conservation relation of $N$ (that is, $c \in L_{\kappa}(N)$, as in Definition~\ref{def:spaces-conservation-laws}), then $c_1\ell_1 + c_2 \ell_2 + \dots + c_n \ell_n$ is an affine-linear steady-state relation of $\widetilde{N}$.  Therefore, we obtain the following $\mathbb{Q}(\kappa)$-linear map:   
    \begin{align*}
    \phi: ~L_{\kappa}(N) ~&\to~ A_{{\kappa}}(\widetilde{N})
    \\
    \notag
    c ~&\mapsto~ c_1 \ell_1 + c_2 \ell_2 + \dots + c_n \ell_n ~.
\end{align*}
\item \label{it:kernel} The  kernel of $\phi$ consists of the linear conservation relations of $N$ that extend to $\widetilde N$: 
    \begin{align*}
    \ker (\phi) &~=~ \{ c \in L_{\kappa}(N) \mid c \in L_{\widetilde{\kappa}}(\widetilde{N}) \}\\
    & ~=~
    \{ c \in L_{\kappa}(N) \mid c_1f_1^{\widetilde{N}} + c_2 f_2^{\widetilde{N}} + \dots + c_nf_n^{\widetilde{N}} = 0\}~
    .\notag
    \end{align*}
  In particular, $\ker (\phi)~ \subseteq~ L_{\kappa}(\widetilde{N})~ \subseteq~ L_{\widetilde{\kappa}}(\widetilde{N})$.

 \item \label{it:all} 
    If, additionally, $N$ is obtained from $\widetilde{N}$ by removing   \uline{all} zero-order and first-order reactions, then the following hold:
    
    \begin{enumerate}[label=(\roman*)]
    \item  \label{it:kernel-if-remove-all}
  The kernel of $\phi$  equals  
    \[
    \ker(\phi) ~=~
         \{c \in \mathbb{Q}(\kappa)^n \mid c_1f^{\widetilde{N}}_1+ c_2f^{\widetilde{N}}_2 + \dots + c_nf^{\widetilde{N}}_n =0\}~,
    \] 
    that is, ${\rm ker}(\phi)= L_{\kappa}(\widetilde{N})$.
    \item \label{it:surjective}
    The map $\phi$ is surjective, that is, ${\rm im}(\phi) =  A_{{\kappa}}(\widetilde{N})$.
       \end{enumerate}
    
     \end{enumerate}

\end{theorem}

\smallskip

The next result follows readily from item~(\ref{it:kernel}) of
Theorem~\ref{thm:main}.

\begin{corollary}[When all linear conservation relations of $N$ extend to $\widetilde{N}$]
    \label{cor:map-is-zero}
    Assume that $N$ is a reaction network obtained from a reaction network $\widetilde{N}$ by removing one or more zero-order  and/or first-order reactions.  The map $\phi$ is the zero map if and only if every linear conservation relation of $N$ is also a linear conservation relation of $\widetilde{N}$.
\end{corollary}

\smallskip

The following result is a direct consequence of the Rank-Nullity Theorem and item~\eqref{it:all} in Theorem~\ref{thm:main}.

\begin{corollary}[Dimension formula]
    \label{coro:dimensionsum}
    Assume that 
    $N$ is a reaction network obtained from a reaction network $\widetilde{N}$ by removing \uline{all} zero-order and first-order reactions.
    Let $\kappa$ and $\widetilde{\kappa}$ denote the vectors of unknown rate constants for $N$ and $\widetilde{N}$, respectively. Then,
\begin{align}
    \label{eq:cor-dim-equation}
\dim_{\mathbb{Q}(\kappa)}(L_{\kappa}(N)) \, = \, 
\dim_{\mathbb{Q}({\kappa})} (A_{{\kappa}}(\widetilde{N}))
~+~ \dim_{\mathbb{Q}(\kappa)}(L_\kappa(\widetilde{N}) )~.
\end{align}
\end{corollary}

\begin{example}[Example~\ref{ex:futile-cycle} continued] \label{ex:1-site}
We revisit the futile cycle $\widetilde{N}$ and its subnetwork $N$ obtained by removing all zero-order and first-order reactions.
We examine the map $\phi: L_{\kappa}(N) \to  A_{{\kappa}}(\widetilde{N})$ from  Theorem~\ref{thm:main}. Recall that $L_{\kappa}(N)$ is 4-dimensional, with a basis given in~\eqref{eq:lin-cons-law-space-1-site}.  We evaluate $\phi$ at each of these basis vectors (and simplify, using~\eqref{eq:conservation-law-subnetwork-of-1-site}) to obtain:
\begin{align*} 
    \phi(e_{S_0} + e_{Y_1})~=~h~, \quad
    \phi(e_{S_1} + e_{Y_2} )  ~=~-h~, \quad
    \phi(e_{E} + e_{Y_1} )  ~=~0~, \quad
    \phi(e_{F} + e_{Y_2} )  ~=~0~,
\end{align*}
where, as before, $h:=k_4[Y_2]-k_2[Y_1]$.  Thus, by Theorem~\ref{thm:main}, the space $A_{{\kappa}}(\widetilde{N})$ is generated by $h$ and hence is 1-dimensional.

Next, as  $L_{\kappa}(N)$ is 4-dimensional and $A_{{\kappa}}(\widetilde{N})$ is 1-dimensional, the kernel of $\phi$ (which equals $L_{\kappa}(\widetilde{N})$ by Theorem~\ref{thm:main}) is $3$-dimensional (as in Corollary~\ref{coro:dimensionsum}).  Recall from Theorem~\ref{thm:main} that this kernel consists of linear conservation relations of $N$ that extend to $\widetilde{N}$.  It is straightforward to check that a basis of such relations comes from the total amounts of substrate $S$, kinase $E$, and phosphatase $F$:
\begin{align}
    \label{eq:futile-cycle-basis-kernel}
    e_{S_0} + e_{Y_1} + e_{S_1} + e_{Y_2}~,\quad 
    e_{E} + e_{Y_1}~, \quad
    e_{F} + e_{Y_2}~.
\end{align}
\end{example}

\bigskip

Recall from Theorem~\ref{thm:main} that, in the context of Setup~\ref{setup:star}, if $N$ has no zero-order or first-order reactions, then the image of $\phi$ equals $A_{{\kappa}}(\widetilde{N})$.  The next result states that, in this setting, the image also generates the larger space $A_{\widetilde{\kappa}}(\widetilde{N})$ as a vector space over $\mathbb{Q}(\widetilde{\kappa})$. 

\begin{proposition}[$A_{{\kappa}}(\widetilde{N})$ spans $A_{\widetilde{\kappa}}(\widetilde{N})$]  
    \label{prop:generates}
If $N$ is a reaction network obtained from a reaction network $\widetilde{N}$ by removing \uline{all} zero-order and/or first-order reactions, then 
 $A_{{\kappa}}(\widetilde{N})$ generates $A_{\widetilde{\kappa}}(\widetilde{N})$ as a vector space over $\mathbb{Q}(\widetilde{\kappa})$.
\end{proposition}
\begin{remark}[Converse of Proposition~\ref{prop:generates} is false] \label{rem:converse-generates}
Example~\ref{ex:eissing-linear} in Section~\ref{sec:Eissing-model} presents  a subnetwork $N$ (of a network $\widetilde{N}$) that contains a first-order reaction, and yet $A_{{\kappa}}(\widetilde{N})$ generates $A_{\widetilde{\kappa}}(\widetilde{N})$ as a vector space over $\mathbb{Q}(\widetilde{\kappa})$.
\end{remark}

\medskip

The next two results in this section (Proposition~\ref{prop:syzygy} and Lemma~\ref{lem:positive_rel} below) allow us to connect steady-state relations to steady states.  The first result is quite general, pertaining to the Jacobian matrix of polynomials having a nontrivial {\em syzygy} (relation), and will be useful for understanding the dimension of the set of steady states (see Section~\ref{sec:Albeck}). The idea is as follows.  
Let $f_1, f_2, \dots, f_m$ denote polynomials in variables $x_1, x_2,\dots, x_n$. If there is a linear relation among these polynomials, $c_1f_1 + c_2f_2 + \dots + c_mf_m=0$, in which the coefficients $c_i$ do {\em not} depend on the variables $x_1,x_2,\dots, x_n$, then this relation clearly generates a linear dependence among the gradients~$\nabla f_i$, that is, among the rows of the Jacobian matrix $J_f = (\frac {\partial f_i} {\partial x_j})_{ij}$.  If, on the other hand, the coefficients $c_i$ depend on the variables $x_1,x_2,\dots, x_n$, then the rows of the Jacobian matrix need not be linearly dependent in general.  However, as the following result asserts, a non-trivial linear dependence can arise from certain syzygies when the Jacobian matrix is evaluated at a common zero $x^*$ of $f_1,f_2,\dots, f_m$.

\begin{proposition}[Rank-deficient Jacobian]\label{prop:syzygy}
Let $f_1, f_2, \dots, f_m \in \mathbb Q(\kappa)[x_1,x_2, \dots, x_n]$, where
$\kappa:=(k_1,k_2,\dots, k_r)$.  
Assume that $\sum_{i=1}^m g_i f_i=0$, for some $g_1,g_2, \dots, g_m \in \mathbb Q(\kappa)[x_1,x_2, \dots, x_n]$. 
Let $\kappa^* \in \mathbb{R}^r_{>0}$
be a vector of positive rate constants  at which the polynomials $g_1,g_2, \dots, g_m$ are defined.
If 
$x^*$ is a common zero of 
$f_1(\kappa^*,x), 
f_2(\kappa^*,x), \dots, f_m(\kappa^*,x)$ but is not a common zero of 
$g_1(\kappa^*,x), 
g_2(\kappa^*,x),
\dots, g_m(\kappa^*,x)$, then
\begin{align}
    \label{eq:dependence-gradient}
\sum_{i=1}^m \, g_i(\kappa^*,x^*) ~\nabla f_i(\kappa^*,x^*) \, = \, {\mathbf 0}
\end{align}
defines a non-trivial linear dependence among the gradients, that is, among the rows of the Jacobian matrix 
of the polynomials $f_i(\kappa^*,x)$
evaluated at $x=x^*$.
    \end{proposition}

\begin{remark}[Syzygies and regular sequences]
The standard notion of linear independence of polynomials (over some field) does not directly translate to the setting of working over the polynomial ring, e.g., $ f_2 \, f_1 + (-f_1) f_2 =0$, for any choice of polynomials $f_1, f_2$. 
In general, the existence of a syzygy $\sum_{i=1}^m g_i f_i$ as in the statement of Proposition~\ref{prop:syzygy} implies that $f_1,f_2, \dots, f_m$ is {\rm not} a {\em regular sequence}.  In turn, the notion of regular sequence is related to the dimension of the zero set (over $\mathbb{C}$) of a finite set of polynomials. 
\end{remark}

The next result pertains to ``positive'' affine-linear steady-state relations, such as $h=x_3$ or $h={\frac{\kappa_3}{\kappa_7}\, x_8+x_{18}}$ (these specific relations appear later in the Albeck {\em et al.}\ model; see Table~\ref{tab:lifted}).  When such relations exist, certain species must be $0$ in  steady state (e.g. $x_3=0$ or $x_8=x_{18}=0$), and thus all non-negative steady states are boundary steady states.  We summarize this idea in the following useful result.

\begin{lemma}[Boundary steady states]\label{lem:positive_rel}
    Let $N$ be a reaction network with $n$ species and $r$ reactions.
    Consider an affine-linear steady-state relation of $N$ of the following form: $h= a_1x_{i_1} + a_2 x_{i_2} + \dots + a_m x_{i_m}$, for some $a_1,a_2, \dots, a_m \in \mathbb{Q}(\kappa)$, where $1 \leq i_1 < i_2< \dots < i_m \leq n$ and $m \geq 1$. 
    
    Assume that, for all $i=1,2,\dots,m$ and for all $\kappa^* \in \mathbb{R}^{r}_{>0}$ in the domain of all $a_1,a_2,\dots,a_m$, we have that $a_i(\kappa^*)>0$.  If $x^* \in \mathbb{R}_{\geq 0}^{n}$ is a steady state of the mass-action system defined by the network $N$ and   the vector of rate constants $\kappa^* \in \mathbb{R}^{r}_{>0}$, then $x^*_{i_1}=x^*_{i_2}=\dots = x^*_{i_m}=0$ and therefore $x^*$ is a boundary steady state.
\end{lemma}

\medskip

We now return to the context of Proposition~\ref{prop:generates}, which we recall gives conditions under which the space $A_{{\kappa}}(\widetilde{N})$, which is a vector space over $\mathbb Q(\kappa)$ associated to a network $N$, generates  
 the space
$A_{\widetilde{\kappa}}(\widetilde{N})$ 
over the larger field $\mathbb Q(\widetilde{\kappa})$.  This second space, which is associated to the larger network $\widetilde{N}$, is more natural to compute, so we would like to know whether a basis for the smaller space, $A_{{\kappa}}(\widetilde{N})$, is in fact also a basis of the larger space, $A_{\widetilde{\kappa}}(\widetilde{N})$.  In other words, is the 
dimension of $A_{{\kappa}}(\widetilde{N})$ over 
$\mathbb Q(\kappa)$ equal to the dimension of 
$A_{\widetilde{\kappa}}(\widetilde{N})$ over $\mathbb Q(\widetilde{\kappa})$?
 
It turns out that these dimensions may differ (as the following example shows).
Accordingly, we define easy-to-check conditions (Hypotheses~\ref{hyp:1} and~\ref{hyp:2} below) that ensure 
that these dimensions coincide (Proposition~\ref{prop:Hi-plus}).

\begin{example}\label{ex:different}
Consider the network $\widetilde{N}=\{X_1+X_2\xrightarrow{k_1}X_3,\ 0\xrightarrow{k_2}X_1,\ 0\xrightarrow{k_3}X_2 \}$. 
By removing all zero-order and first-order reactions, we obtain $N=\{X_1+X_2\xrightarrow{k_1}X_3 \}$, with conservation laws arising from the equalities $f_1^N+f_3^N=0$ and $f_2^N+f_3^N=0$. 
It is now straightforward to apply Theorem~\ref{thm:main} to conclude that
$A_{{\kappa}}(\widetilde{N})$ is generated by two constant functions, $\{k_2, k_3\}$. This set $\{k_2, k_3\}$ is linearly independent over $\mathbb Q(\kappa) = \mathbb Q(k_1)$, but is linearly dependent over  
$\mathbb Q(\widetilde{\kappa}) = \mathbb Q(k_1, k_2, k_3)$.
Indeed, for this network,
\[
\dim_{\mathbb Q (\kappa)}(A_\kappa(\widetilde{N})) ~=~ 2 ~>~ 1 ~=~ \dim_{\mathbb Q(\widetilde{\kappa})}(A_{\widetilde{\kappa}}(\widetilde{N}) )~.
\]
\end{example}

In what follows, we again let $N$ denote a subnetwork 
obtained from a network $\widetilde{N}$ by removing all zero-order and first-order reactions.   For convenience, we denote the stoichiometric subspaces of $N$ and $\widetilde{N}$, as in~\eqref{eq:defS}, by (respectively) $S_N$ and $S_{\widetilde{N}}$.

\begin{notation} \label{notation:zero-cpx}
    We use $X_0$ to denote the zero complex: $X_0:=0$.  Also, a {\em reaction out of~$X_j$} simply means a reaction for which the reactant complex is $X_j$ (such as $X_j \to 0$).  Finally, a {\em removed} reaction is one that is in $\widetilde{N}$, but not in $N$.
\end{notation}

\begin{hypothesis} \label{hyp:1}
For each $j=0,1,\dots,n$, at most one removed reaction out of $X_j$ satisfies the following: its reaction vector (which lies in the stoichiometric subspace $S_{\widetilde{N}}$ of $\widetilde{N}$) does not lie in the stoichiometric subspace $S_N$ of the smaller network $N$. Moreover, for all $c \in \mathbb Q(\kappa)^n$ such that $\sum_{i=1}^n c_i f_i^N=0$ (that is, $c \in L_\kappa(N)$), the vector $c$ is  orthogonal to all reaction vectors of $N$.
\end{hypothesis}

\begin{remark}
 \label{rem:hyp-1-part-2}   
The second part of Hypothesis~\ref{hyp:1} is satisfied when each linkage class  (connected component of the reaction graph) of $N$ has exactly one terminal strong linkage class~\cite{feinberg-horn-1977}. This graphical condition is easily checked. 
\end{remark}

\begin{example}[Example~\ref{ex:1-site} continued]
    \label{ex:1-site-hypothesis}
We show that Hypothesis~\ref{hyp:1} holds for the futile cycle.  
Recall the removed reactions:
        $S_0+E  \leftarrow Y_1 \rightarrow S_1 +E$ and 
        $S_1+F   \leftarrow Y_2 \rightarrow S_0+F$. 
The reactions
$S_0+E  \leftarrow Y_1$
and 
$S_1+F   \leftarrow Y_2$
are the reverse reactions of the reactions in~$N$, so their reaction vectors belong to $S_N$. 
Additionally, each of the two linkage classes of the network $N=\{S_0+E\to Y_1,\ S_1+F\to Y_2\}$ has only one terminal strong linkage class, as in Remark~\ref{rem:hyp-1-part-2}.
Hypothesis~\ref{hyp:1} therefore holds for this example.
\end{example}

\medskip

In later sections, we analyze two
models of caspase activation in apoptosis, the Eissing {\em et al.}\ model and the Albeck {\em et al.} model.  
We will see that the Albeck \textit{et al.}\ model satisfies Hypothesis~\ref{hyp:1} (see Section~\ref{sec:Albeck}). However, this hypothesis does not hold for the Eissing \textit{et al.} model, due to the four removed reactions that are inflow (synthesis) reactions (see Figure~\ref{fig:eissing-model} and Table~\ref{tab:eissing-reactions}). For this model, we need to introduce another sufficient hypothesis.

\begin{hypothesis}\label{hyp:2}
 For some choice of a basis $\{d^1,d^2,\dots, d^s\}$ for $L_\kappa(N)$  over $\mathbb{Q}(\kappa)$,  there exist distinct species $X_{j_1}, X_{j_2},\dots,X_{j_s}$ such that, for each $1\leq \ell \leq s$, the following hold:
\begin{enumerate}[label=(\alph*)]
\item the only removed reaction out of $X_{j_{\ell}}$ is the degradation reaction $X_{j_{\ell}} \to 0$;
\item 
$d^{\ell}_{j_{\ell}}\neq0$ and 
$d^{\ell}_{j_{\ell+1}}=d^{\ell}_{j_{\ell+2}}=\dots =d^{\ell}_{j_{s}}=0$
(that is, the support of $d^{\ell}$ contains $j_{\ell}$, but not any of $j_{\ell+1},j_{\ell+2},\dots,j_s$). 
\end{enumerate}
\end{hypothesis}

\noindent
In Section~\ref{sec:Eissing-model}, 
we will see that the Eissing {\em et al.}\ model satisfies Hypothesis~\ref{hyp:2} (see Example~\ref{ex:eissing-follow-up-on-intro}).

\medskip

We can now state the following result:

\begin{proposition}\label{prop:Hi-plus}

Let $N$ be a reaction network obtained from a reaction network $\widetilde{N}$ by removing   \uline{all} zero-order and first-order reactions.
If  Hypothesis~\ref{hyp:1} or Hypothesis~\ref{hyp:2} holds, then the following equality holds:
\begin{equation}\label{eq:kk}
 \dim_{\mathbb{Q}(\kappa)}(A_{\kappa}(\widetilde{N}))~ =~ \dim_{\mathbb{Q}(\widetilde{\kappa})}(A_{\widetilde{\kappa}}(\widetilde{N}))~.
\end{equation}

\end{proposition}

\medskip

In the next sections, we apply our above results
to the models of caspase activation mentioned earlier, the Eissing {\em et al.}\ model (Section~\ref{sec:Eissing-model}) and the Albeck {\em et al.} model (Section~\ref{sec:Albeck}).  
Some of our computations are supported by supplementary {\tt Maple} files~\cite{maple18}.
A summary of examples arising from these models -- and the futile cycle analyzed earlier -- is in Table~\ref{tab:summary-examples}.

\begin{table}[ht]
\begin{tabular}{lcc}
\hline
~
\\
Example    & Does ${\rm im}(\phi)$ generate $A_{\widetilde{\kappa}}(\widetilde{N})$?\\
\hline
Futile cycle (Ex.~\ref{ex:1-site})     
& Yes \\                  
Eissing, all 0/1-order removed (Ex.~\ref{ex:eissing-follow-up-on-intro}) & 
Yes 
\\
Eissing, all but $k_{-11}$ removed (Ex.~\ref{ex:eissing-linear})    
    & Yes                         \\
Eissing, all but $k_{-3}$ removed (Ex.~\ref{ex:eissing-keep-ell-3}) & 
No 
\\
Albeck, EARM (Sec.~\ref{sec:Albeck})        & Yes                         \\
\hline
\end{tabular}

    \caption{Summary of applications of Theorem~\ref{thm:main}: examples of networks and whether the image of the map $\phi$ generates $A_{\widetilde{\kappa}}(\widetilde{N})$ as a $\mathbb{Q}(\widetilde{\kappa})$-vector space.
    }
    \label{tab:summary-examples}
\end{table}

\medskip

We end this section with an example where no (nonzero) linear relations at steady state can be obtained with coefficients only depending on the rate constants.

\begin{example}
    \label{ex:one-target}
    Consider the following network $N$, which has $3$ species and $3$ reactions:
    
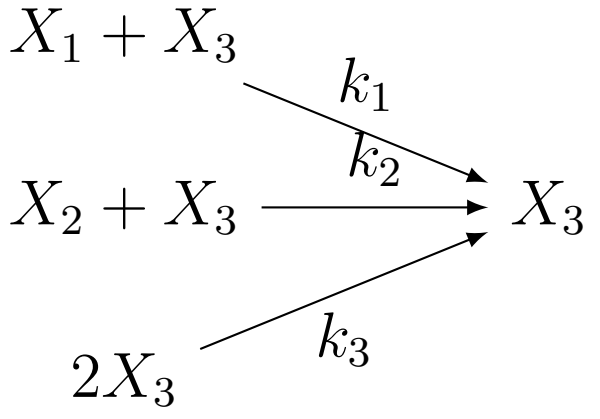
\begin{center}
\begin{tikzpicture}[node distance=0.5cm and 1.5cm,scale=0.7]

\node (13) {$X_1+X_3$};
\node (23) [below= of 13] {$X_2+X_3$};
\node (33) [below= of 23] {$2X_3$};
\node (3) [right=of 23] {$X_3$};

\draw[->,> = latex] (13) -- (3) node[midway, above] {$k_1$};
\draw[->,> = latex] (23) -- (3) node[midway, above] {$k_2$};
\draw[->,> = latex] (33) -- (3) node[midway, below] {$k_3$};
\end{tikzpicture}
\end{center}
Note that $X_3$ is the unique product complex.  
The resulting mass-action ODEs are as follows:
    \begin{align*}
        \frac{dx_1}{dt} ~&=~  - k_1 \, x_1 x_3 ~=:~ f_1^N, \quad \quad 
        \frac{dx_2}{dt} ~=~ - k_2 \, x_2 x_3  ~=:~ f_2^N, \quad \quad
        \frac{dx_3}{dt} ~=~  - k_3 \, x_3^2 ~=:~ f_3^N~.
    \end{align*}
The monomials appearing in the $f_i^N$ are distinct, so $N$ has no linear conservation relations ($L_{\kappa}(N)=\{0\}$).  Next, consider the network $\widetilde{N}$ obtained from $N$ by adding inflow
reactions:
\begin{align*}
    0
    \xrightarrow{\gamma_1} X_1~,
    \quad 
    0
    \xrightarrow{\gamma_2} X_2~,
    \quad 
    0 
    \xrightarrow{\gamma_3} X_3    ~,
\end{align*}
which results in the following mass-action ODEs~\eqref{eq:mass_action_ODE} for $\widetilde{N}$: 
    \begin{align*}
        \frac{dx_1}{dt} ~&=~  f_1^N +  \gamma_1  ~=:~ f_1^{\widetilde{N}}, \quad \quad 
        \frac{dx_2}{dt} ~=~ f_2^N +  \gamma_2 ~=:~ f_2^{\widetilde{N}}, \quad \quad 
        \frac{dx_3}{dt} ~=~ f_3^N +  \gamma_3 ~~=:~ f_3^{\widetilde{N}}~.
    \end{align*}
Consider the following polynomial:
$h:=
 (k_2 \, x_2) \, f_1^{\widetilde{N}} - (k_1 x_1)\, f_2^{\widetilde{N}} \, = \, k_2 \gamma_1 \, x_2 - k_1 \gamma_2 \, x_1$,
which is affine linear (in the variables $x_i$).  
As $h$ is a polynomial-combination of $f_1^{\widetilde{N}}$ and $f_2^{\widetilde{N}}$, we conclude that $h$ vanishes at steady state.  Hence, $h$ is an affine-linear steady-state relation of $\widetilde{N}$.  However, $h$ can \uline{not} be obtained {\em linearly} (i.e., with constant coefficients, as opposed to {\em polynomially}) from the polynomials $f_i^{\widetilde{N}}$; this fact can be seen by examining the monomials appearing in the $f_i^{\widetilde{N}}$. In fact, it is straightforward to check that no (nonzero) affine-linear steady-state relations are obtained {\em linearly}, i.e., $A_{{\kappa}}(\widetilde{N}) =A_{\widetilde{\kappa}}(\widetilde{N})=\{0\}$, where $\kappa=(k_1,k_2, k_3)$ and
$\widetilde{\kappa} =(k_1,k_2,k_3,\gamma_1,\gamma_2,\gamma_3)$
denote, as usual,  the vectors of unknown rate constants of $N$ and $\widetilde{N}$, respectively.
\end{example}

\section{Eissing {\em et al.}\ model} \label{sec:Eissing-model}

In this section, we return to the Eissing {\em et al.}\ model of caspase activation (Figure~\ref{fig:eissing-model}), with ODEs given earlier, in~\eqref{eq:Eissing-ODE}.  We consider three subnetworks. The first is obtained by removing all zero-order and first-order reactions (this is the example described in the introduction, with more details given below in Example~\ref{ex:eissing-follow-up-on-intro}).  In contrast, the subnetworks in Examples~\ref{ex:eissing-linear} and~\ref{ex:eissing-keep-ell-3} contain a first-order reaction.

\begin{example}[Eissing {\em et al.}\ model, all 0/1-order reactions removed]
\label{ex:eissing-follow-up-on-intro}
Let $\widetilde{N}$ denote the network in Figure~\ref{fig:eissing-model}, which underlies the Eissing {\em et al.}\ model.  The vector of rate constants is $\widetilde{\kappa}:=(k_1,k_2,\dots, k_{13}, ~k_{-3},k_{-8},k_{-9},k_{-10}, k_{-11},k_{-12})$. 
We return to the subnetwork~\eqref{eq:Eissing-subnetwork-smallest}, which we call $N$, which is obtained from $\widetilde{N}$ by removing all zero-order and first-order reactions.  
We recast the $4$ (linearly independent) conservation relations of $N$, in~\eqref{eq:Eissing-smallest-subnetwork-cons-laws}, as follows: 
    \begin{align} \label{eq:lin-cons-law-space-Eissing-subnetwork-smallest}
        L_{\kappa}(N) ~& \supseteq ~ \mathrm{span}_{\mathbb{Q}(\kappa)} \left( \{ 
        e_1 + e_2 + e_8,~
        e_3 + e_4 + e_6,~
        e_7 + e_8 ,~
        k_3 e_5 + (k_3+k_4) e_6         
         \} \right)~,
    \end{align}
where $e_1, e_2,\dots, e_8$ denote the canonical basis vectors and 
$\kappa:=(k_1, k_2, k_3, k_4, k_{11})$. 
The containment~\eqref{eq:lin-cons-law-space-Eissing-subnetwork-smallest} is, in fact, an equality (see supplementary {\tt Maple} file) and thus $L_{\kappa}(N)$ is $4$-dimensional.

Next, consider the map $\phi: L_{\kappa}(N) \to A_{{\kappa}}(\widetilde{N})$ from Theorem~\ref{thm:main}. 
By evaluating $\phi$ at each of the $4$ basis vectors shown on the right-hand side of~\eqref{eq:lin-cons-law-space-Eissing-subnetwork-smallest}, and then simplifying, we get:
\begin{align} \label{eq:eissing-affine-linear}
        \notag
        \phi ( e_1 + e_2 + e_8) ~&=~ - k_9 x_1 - k_5 x_2 -k_{13} x_8 + k_{-9}~ =:~ h_1~,~
        \\
        \phi (e_3 + e_4 + e_6) ~&=~ -k_{10} x_3 - k_6 x_4 - k_7 x_6 + k_{-10}~ =:~ h_2~,~
        \\
        \notag
        \phi (e_7 + e_8) ~&=~ -k_{12} x_7 - k_{13} x_8 + k_{-12} ~ =:~ h_3~,~
        \\
        \notag
        \phi (k_3 e_5 + (k_3+k_4) e_6) ~&=~  -k_3 k_8 x_5 -  (k_3 k_7 + k_4 k_7 + k_4 k_{-3})x_6 + k_{-8} k_3~ =:~ h_4~.
\end{align}
The expressions $h_1,h_2,h_3,h_4$ are affine-linear steady-state relations for the original network~$\widetilde{N}$; and they exactly match the ones shown earlier, in \eqref{eq:eissing-4-linear0}.  Additionally, $\{ h_1, h_2, h_3 ,h_4\}$ is easily seen to be a linearly independent set over $\mathbb{Q}(\widetilde{\kappa})$ (for details, see the end of this example).  Thus, as $\phi$ is surjective (by part~(ii) of Theorem~\ref{thm:main})
and its image generates $A_{\widetilde{\kappa}}(\widetilde{N})$ (by Proposition~\ref{prop:generates}), we conclude that $\{ h_1, h_2, h_3 ,h_4\}$ is a basis of the space $A_{\widetilde{\kappa}}(\widetilde{N})$ of affine-linear steady state relations.

 We now give an another approach to showing that the space $A_{\widetilde{\kappa}}(\widetilde{N})$ is $4$-dimensional.  Specifically, we will see that this example satisfies Hypothesis~\ref{hyp:2}, which then, by Proposition~\ref{prop:Hi-plus}, ensures that $\dim_{\mathbb{Q}(\kappa)}(A_{\kappa}(\widetilde{N}))=\dim_{\mathbb{Q}(\widetilde{\kappa})}(A_{\widetilde{\kappa}}(\widetilde{N}))$.  Indeed, our computations below capture some main ideas behind the proof of Proposition~\ref{prop:Hi-plus}, which appears in Section~\ref{sec:proofs}.

To verify Hypothesis~\ref{hyp:2}, we return to the $4$ basis vectors of $L_{\kappa}(N)$ on the right-hand side of~\eqref{eq:lin-cons-law-space-Eissing-subnetwork-smallest}.  We label them by $d^1:=e_1 + e_2 + e_8$, $d^2:=e_3 + e_4 + e_6$, $d^3:=e_7 + e_8$, and $d^4:=k_3 e_5 + (k_3+k_4) e_6$. 
Consider $j_1=2$, $j_2=4$, $j_3=7$, and $j_4=5$. 
We see from Table~\ref{tab:eissing-reactions} that the only reaction out of $X_{j_{\ell}}$ is its degradation reaction ($X_2\overset{k_5}{\longrightarrow}0$, $X_4\overset{k_6}{\longrightarrow}0$, $X_7\overset{k_{12}}{\longrightarrow}0$, $X_5\overset{k_8}{\longrightarrow}0$); and these reactions are removed when going from $\widetilde{N}$ to $N$.  Also, $e_2$ (respectively, $e_4,e_7,e_5$) appears only in $d^1$ (respectively, $d^2,d^3,d^4$).  Thus, Hypothesis~\ref{hyp:2} holds.

Our next aim is to show how the properties of the $d^i$ ensure that $h_1,h_2,h_3,h_4$ from~\eqref{eq:eissing-affine-linear} are linearly independent over $\mathbb{Q}(\widetilde{\kappa})$ (and thus, as explained earlier, form a $\mathbb{Q}(\widetilde{\kappa})$-basis).  We begin with the
matrix of coefficients of $h_1,h_2,h_3,h_4$ (from~\eqref{eq:eissing-affine-linear}), where the ordering of the variables $x_i$ is as shown (and the last column corresponds to the constant term, via $x_0:=1$):
\[
\begin{bNiceMatrix}[first-row]
x_2 & x_4 & x_7 & x_5 &x_1 &x_3 &x_6 &x_8 &x_0  \\ 
 - k_5 & 0 & 0 & 0 & - k_9 & 0 & 0 & -k_{13} & k_{-9} \\
  0 & - k_6 & 0 & 0 & 0 & -k_{10} & - k_7 & 0 & k_{-10}\\
  0 & 0 &  -k_{12} & 0 & 0 & 0 & 0 & - k_{13} & k_{-12} \\
  0 & 0 & 0 &  -k_3 k_8 & 0 & 0 &  -  (k_3 k_7 + k_4 k_7 + k_4 k_{-3}) & 0 &  k_{-8} k_3  \end{bNiceMatrix}~.
\]

By inspecting the first $4$ columns of this coefficient matrix, we easily deduce that the $4$ rows of the matrix are linearly independent over  $\mathbb{Q}(\widetilde{\kappa})$ and thus $\{h_1,h_2,h_3,h_4\}$ is linearly independent over $\mathbb{Q}(\widetilde{\kappa})$. 
We conclude that the equality  $\dim_{\mathbb{Q}(\kappa)}(A_{\kappa}(\widetilde{N}))=\dim_{\mathbb{Q}(\widetilde{\kappa})}(A_{\widetilde{\kappa}}(\widetilde{N}))$, from Proposition~\ref{prop:Hi-plus}, holds.
We end this example by illustrating a key idea in the proof of Proposition~\ref{prop:Hi-plus}: 
the $4\times 4$ submatrix corresponding to the first $4$ columns of the coefficient matrix (shown earlier) can be factored as a product of two diagonal matrices: 
    \[
    \begin{pmatrix}
        -1 & & & \\
        & -1 & & \\
        & & -1 & \\
        & & & -k_3
    \end{pmatrix}\begin{pmatrix}
        k_5 & & & \\
        & k_6 & & \\
        & & k_{12} & \\
        & & & k_8
    \end{pmatrix}.
    \]
Observe that the matrix on the left involves entries in $\mathbb{Q}(\kappa)=\mathbb{Q}(k_1, k_2, k_3, k_4, k_{11})$. In contrast, the matrix on the right involves the rate constants $k_5$, $k_6$, $k_8$, and $k_{12}$, which correspond to the removed reactions, and therefore the entries lie in the larger field $\mathbb{Q}(\widetilde{\kappa})$.  Additionally, both matrices are invertible for any choice of positive rate constants $k_i$.

\end{example}

\begin{remark}[Linear steady-state relations in Carotenuto {\em et al.}~\cite{analyse-eissing}] \label{rem:linear-carotenuto}
    Three of the affine-linear steady state relations in~\eqref{eq:eissing-affine-linear} -- namely, $h_1$, $h_2$, and $h_3$ -- appear as equations~(2.10), (2.12), and~(2.11), respectively, in the steady-state analysis of the Eissing {\em et al.}\ model performed by Carotenuto {\em et al.}~\cite{analyse-eissing}.  
\end{remark}

\medskip

\begin{example}[Eissing {\em et al.} model, all 0/1-order reactions removed except $X_8 \to X_2+X_7$] \label{ex:eissing-linear}
        We again let $\widetilde{N}$ denote the full Eissing~{\em et al.}\ network in Figure~\ref{fig:eissing-model}, but this time we consider the following subnetwork $\widehat{N}$ obtained by removing all zero-order and first-order reactions, except the reaction labeled below by $k_{-11}$:
    \begin{align*}
        X_1+X_4 \xrightarrow{k_2}  
        X_2 + X_4 \xleftarrow{k_1} X_2+X_3
        \quad 
        X_4 \xleftarrow{k_4} X_4+X_5 \xrightarrow{k_3} X_6 \quad 
        X_2 + X_7 \stackrel[k_{-11}]{k_{11}}{\rightleftarrows} X_8~.  
    \end{align*}       
    Accordingly, let $\widehat{\kappa}:=(k_1, k_2, k_3, k_4, k_{11},k_{-11})$. 
    
    The analysis of this example is remarkably similar to that of the smaller network $N$, from Example~\ref{ex:eissing-follow-up-on-intro}.  To begin with, 
    the linear conservation relations of $N$ are also linear conservation relations of $\widehat{N}$; in other words, the containment~\eqref{eq:lin-cons-law-space-Eissing-subnetwork-smallest} holds, with $N$ and $\kappa$ replaced by, respectively, $\widehat{N}$ and $\widehat{\kappa}$.
    (The fourth conservation relation comes from the fact that $f^{\widehat{N}}_5=- k_3 x_4 x_5 -  k_4 x_4 x_5$ and $f^{\widehat{N}}_6 = k_3 x_4 x_5$, where we use the notation $f^{\widehat{N}}_i$ for the right-hand sides of the ODEs of ${\widehat{N}}$, as in Theorem~\ref{thm:main}.) 
    Also, as before, the containment is an equality, and so 
    $L_{\widehat{\kappa}}(\widehat{N})$ is $4$-dimensional (see supplementary {\tt Maple} file).

Next, consider the map $ L_{\widehat{\kappa}}(\widehat{N}) \to A_{\widehat{\kappa}}(\widetilde{N})$ from Theorem~\ref{thm:main}, 
which we denote by $\phi_{\widehat{N}}$ (rather than $\phi$) to avoid confusion with the map from Example~\ref{ex:eissing-follow-up-on-intro}.
When we evaluate $\phi_{\widehat{N}}$ at the $4$ linear conservation relations, we obtain the expressions $h_1,h_2,h_3,h_4$ shown in~\eqref{eq:eissing-affine-linear}. 
We saw, in Example~\ref{ex:eissing-follow-up-on-intro}, that $\{h_1,h_2,h_3,h_4\}$ is a basis of $A_{\widetilde{\kappa}}(\widetilde{N})$, so 
the image of $\phi_{\widehat{N}}$ generates $A_{\widetilde{\kappa}}(\widetilde{N})$ as a vector space over $\mathbb{Q}(\widetilde{\kappa})$.
Thus, as ${\rm im}(\phi_{\widehat{N}}) \subseteq A_{\widehat{\kappa}}(\widetilde{N})$, we conclude that $A_{\widehat{\kappa}}(\widetilde{N})$ also generates $A_{\widetilde{\kappa}}(\widetilde{N})$ (over $\mathbb{Q}(\widetilde{\kappa})$).

    \end{example}

\begin{example}[Eissing {\em et al.} model, all 0/1-order reactions removed except $X_6 \to X_4+X_5$] \label{ex:eissing-keep-ell-3}
As before, let $\widetilde{N}$ denote the full Eissing~{\em et al.}\ network in Figure~\ref{fig:eissing-model}.  We focus on the subnetwork $N_0$ obtained by removing all zero-order and first-order reactions, except  $X_6 \xrightarrow{k_{-3}} X_4+X_5$:
    \begin{align*}
        X_1+X_4 \xrightarrow{k_2}  
        X_2 + X_4 \xleftarrow{k_1} X_2+X_3
        \quad 
        X_4 \xleftarrow{k_4} X_4+X_5 \stackrel[k_{-3}]{k_{3}}{\rightleftarrows} X_6 \quad 
        X_2 + X_7 \xrightarrow{k_{11}} X_8~.  
    \end{align*}       

Unlike in Examples~\ref{ex:eissing-follow-up-on-intro} and~\ref{ex:eissing-linear}, every linkage class has only one terminal strong linkage class (i.e., strongly connected component with no reactions leaving it); the key difference is in the ``middle'' linkage class (containing the complexes $X_4$, $X_4+X_5$, and $X_6$).  A result of Feinberg and Horn~\cite[Corollary~2]{feinberg-horn-1977} now guarantees that the only linear conservation relations of $N_0$ are the three arising from stoichiometry (recall Remark~\ref{rem:stoichiometry}), representing total amounts, as shown earlier in the first three equations in~\eqref{eq:Eissing-smallest-subnetwork-cons-laws}.  

Thus, the image of the resulting map $\phi$, from 
Theorem~\ref{thm:main}, has dimension at most $3$ (as a $\mathbb{Q}(\kappa)$-vector space).  
On the other hand, we saw in Example~\ref{ex:eissing-follow-up-on-intro} that $A_{\widetilde{\kappa}}(\widetilde{N})$ is a $4$-dimensional $\mathbb{Q}(\widetilde{\kappa})$-vector space.  
We conclude that the image of $\phi$ does not generate $A_{\widetilde{\kappa}}(\widetilde{N})$ as a $\mathbb{Q}(\widetilde{\kappa})$-vector space.
\end{example}

\section{Albeck {\em et al.} model (the extrinsic apoptosis reaction model)} \label{sec:Albeck}

In this section, we analyze the ``extrinsic apoptosis reaction model'' (EARM), which was constructed by Albeck {\em et al.}~\cite{albeck2008}.  This model, which arises from the reaction network depicted in Figure~\ref{fig:EARM}, 
consists of four interacting subcircuits~\cite{albeck2008}.  These subcircuits represent the activation of pro-caspase-8, the activation of caspase-3 (which can cleave PARP, the DNA damage repair enzyme), a mitochondrial feed-forward pathway, and a positive feedback loop in which pro-caspase-6 (pro-C6) is cleaved by the active form of caspase-3 to form C6$^*$, which then activates pro-caspase-8~\cite{albeck2008}.  The $70$ reactions in this reaction network involve $58$ species (these species are listed in Table~\ref{tab:albeck-species}).

For this model, we will see that Theorem~\ref{thm:main} yields $15$ affine-linear steady state relations.  
These $15$ relations will allow us, via Lemma~\ref{lem:positive_rel}, to  reduce the number of variables by $17$.  Additional analyses, involving Gr\"obner bases and the linear conservation laws of the original network, will reveal additional features of this network, as summarized in Table~\ref{table:EARM} (see also Theorem~\ref{thm:albeck} at the end of this section).

\begin{table}[ht!]
 \renewcommand{\arraystretch}{2.5}
    \begin{tabular}{|c|c|}
    \hline
    \begin{minipage}{0.4\textwidth} 
    It has $58$ equations in $58$ variables. 
    \end{minipage} & 
    \begin{minipage}{0.45\textwidth} 
    The subnetwork $N$ with all 0/1-order reactions removed has $33$ conservation laws.
    \end{minipage}
    \\
    \hline
    \begin{minipage}{0.4\textwidth} 
    It has $18$ conservation laws. 
    \end{minipage} & 
    \begin{minipage}{0.45\textwidth} 
    It has $15$ linear steady state relations obtained from $\phi(L_\kappa(N))$.
    \end{minipage}
    \\
    \hline
    \begin{minipage}{0.4\textwidth} 
    Every species appears in a conservation law, so the stoichiometric compatibility classes are compact. 
   \smallskip
    \end{minipage} 
     & 
    \begin{minipage}{0.45\textwidth} 
    There are no positive steady states. The 
    set of steady states within a compatibility class can have dimension $2$.
    \smallskip
    \end{minipage}
    \\
    \hline
    \begin{minipage}
 {0.4\textwidth} All steady states are degenerate
    \end{minipage}
   &
    \begin{minipage}
 {0.4\textwidth} The decrease in the rank of the Jacobian at the steady states is not due only to linear conservation laws.

 \vspace{2mm}
    \end{minipage}
    \\
    \hline
    \end{tabular}
 \caption{Main features of the Albeck {\em et al.}\ model.  
}\label{table:EARM}
\end{table}

The remainder of this section is organized as follows.  In Section~\ref{sec:albeck-linear-relation}, we use the affine-linear steady state relations to reduce the number of variables by $17$.  Next, in Section~\ref{sec:albeck-boundary}, we use Gr\"obner bases to further reduce the number of variables, which reveals that the set of steady states is defined by the vanishing of certain monomials and allows us to analyze the degeneracy and dimension of the steady states.
Our computations can be found in a supplementary {\tt Maple} file.

\begin{table}[ht]
    \centering
 {\small
     \begin{tabular}{|@{\hspace{2pt}}c@{\hspace{1pt}}c|@{\hspace{2pt}}c@{\hspace{1pt}}c|@{\hspace{2pt}}c@{\hspace{1pt}}c|@{\hspace{2pt}}c@{\hspace{1pt}}c|@{\hspace{2pt}}c@{\hspace{1pt}}c|@{\hspace{2pt}}c@{\hspace{1pt}}c|}
        \hline
        $x_1$  & [L] & $x_{11}$ & [C8$^*$:Bar] & $x_{21}$ & [PARP] & $x_{31}$ & [Bax$^*$] & $x_{41}$ & [M$^*$] & $x_{51}$ & [Apaf$^*$]\\
        \hline
        $x_2$  & [R] & $x_{12}$ & [C3] & $x_{22}$ & [C3$^*$:PARP] & $x_{32}$ & [Bax$^*_m$] & $x_{42}$ & [CyC$_m$] & $x_{52}$ & [C9]\\
        \hline
        $x_3$  & [L:R] & $x_{13}$ & [C8$^*$:C3] & $x_{23}$ & [cPARP] & $x_{33}$ & [Bcl2] & $x_{43}$ & [M$^*$:CyC$_m$] & $x_{53}$ & [Apop]\\
        \hline
        $x_4$  & [R$^*$] & $x_{14}$ & [C3$^*$] & $x_{24}$ & [Bid] & $x_{34}$ & [Bax$^*_m$:Bcl2] & $x_{44}$ & [CyC$_r$] & $x_{54}$ & [Apop:C3]\\
        \hline
        $x_5$  & [flip] & $x_{15}$ & [C6] & $x_{25}$ & [C8$^*$:Bid] & $x_{35}$ & [Bax$_2$] & $x_{45}$ & [Smac$_m$] & $x_{55}$ & [Smac]\\
        \hline
        $x_6$  & [flip:R$^*$] & $x_{16}$ & [C3$^*$:C6] & $x_{26}$ & [tBid] & $x_{36}$ & [Bax$_2$:Bcl2] & $x_{46}$ & [M$^*$:Smac$_m$] & $x_{56}$ & [Apop:XIAP]\\
        \hline
        $x_7$  & [C8] & $x_{17}$ & [C6$^*$] & $x_{27}$ & [Bcl2$_c$] & $x_{37}$ & [Bax$_4$] & $x_{47}$ & [Smac$_r$] & $x_{57}$ & [Smac:XIAP]\\
        \hline
        $x_8$  & [C8:R$^*$] & $x_{18}$ & [C6$^*$:C8] & $x_{28}$ & [tBid:Bcl2$_c$] & $x_{38}$ & [Bax$_4$:Bcl2] & $x_{48}$ & [CyC] & $x_{58}$ & [C3$^*$$_{Ub}$]\\
        \hline
        $x_9$  & [C8$^*$] & $x_{19}$ & [XIAP] & $x_{29}$ & [Bax] & $x_{39}$ & [M] & $x_{49}$ & [Apaf] &  & \\
        \hline
        $x_{10}$  & [Bar] & $x_{20}$ & [XIAP:C3$^*$] & $x_{30}$ & [tBid:Bax] & $x_{40}$ & [Bax$_4$:M] & $x_{50}$ & [Apaf:CyC] &  & \\
        \hline
     \end{tabular}}
    \caption{Species in the Albeck {\em et al.}\ model.  
    Notably, $x_{12}$ and $x_{7}$ represent
    caspase-3 and pro-caspase-8, respectively; and $x_{14}$ and $x_9$ represent the activated forms.  For details on the remaining species, see Albeck {\em et al.}~\cite{albeck2008}.} 
    \label{tab:albeck-species}
\end{table}

\subsection{Affine-linear steady state relations} \label{sec:albeck-linear-relation}
The Albeck {\em et al.} model is the mass-action kinetics model arising from the reaction network in Figure~\ref{fig:EARM}~\cite{albeck2008}.  Therefore, there are $58$ differential equations in the $58$ species variables $x_1,x_2,\dots, x_{58}$, involving $70$ reaction rate constants.  
The model has the following $18$ linearly independent conservation laws, all of which arise from stoichiometry:  
\begin{align*}
    x_{1}+x_{3}+x_{4}+x_{6}+x_{8}=&T_{1}
    & 
     x_{24}+x_{25}+x_{26}+x_{28}+x_{30}=&T_{10}
\\
x_{2}+x_{3}+x_{4}+x_{6}+x_{8}=&T_{2}
    & 
 x_{27}+x_{28}=&T_{11}
\\
  x_{5}+x_{6}=&T_{3}
    & 
 x_{33}+x_{34}+x_{36}+x_{38}=&T_{13}
\\
 x_{7}+x_{8}+x_{9}+x_{11}+x_{13}+x_{18}+x_{25}=&T_{4}
    & 
 x_{39}+x_{40}+x_{41}+x_{43}+x_{46}=&T_{14}
\\
 x_{10}+x_{11}=&T_{5}
    &
 x_{42}+x_{43}+x_{44}+x_{48}+x_{50}=&T_{15}
\\
 x_{12}+x_{13}+x_{14}+x_{16}+x_{20}+x_{22}+x_{54}+x_{58}=&T_{6}
    &
 x_{45}+x_{46}+x_{47}+x_{55}+x_{57}=&T_{16}
\\
 x_{15}+x_{16}+x_{17}+x_{18}=&T_{7}
 &
 x_{49}+x_{50}+x_{51}+x_{53}+x_{54}+x_{56}=&T_{17}
\\
  x_{19}+x_{20}+x_{56}+x_{57}=&T_{8}
  &
  x_{52}+x_{53}+x_{54}+x_{56}=&T_{18}~
\\
x_{21}+x_{22}+x_{23}=&T_{9}
&
\end{align*}
\vspace{-0.3in}
\begin{align*}    x_{29}+x_{30}+x_{31}+x_{32}+x_{34}+2x_{35}+2x_{36}+4x_{37}+4x_{38}+4x_{40}+4x_{41}+4x_{43}+4x_{46}=&T_{12}
\end{align*}
Therefore, the original $58$ steady-state equations can be reduced to $40$ linearly independent equations in $40$ variables.

Next, observe that each species variable $x_i$ appears in at least one conservation law (with positive coefficient).  Hence, every stoichiometric compatibility class is compact.
A version of the Brouwer fixed-point theorem
guarantees that, for any choice of positive rate constants, every stoichiometric compatibility class contains at least one non-negative steady state.

\begin{figure}
 \begin{adjustbox}{max width=\textwidth}
\begin{tikzcd}[column sep=15pt,row sep=large]
X_1+X_2
  \arrow[r,shift left=.8ex,"k_1"]
&
X_3
  \arrow[l,shift left=.8ex,color=red,"k_{-1}"]
  \arrow[r,color=red,"\kappa_1"]
&
X_4
&
X_4+X_5
  \arrow[r,shift left=.8ex,"k_2"]
&
X_6
  \arrow[l,shift left=.8ex,color=red,"k_{-2}"]
&
X_4+X_7
  \arrow[r,shift left=.8ex,"k_3"]
&
X_8
  \arrow[l,shift left=.8ex,color=red,"k_{-3}"]
  \arrow[r,color=red,"\kappa_3"]
&
X_4+X_9
\\
X_9+X_{10}
  \arrow[r,shift left=.8ex,"k_4"]
&
X_{11}
  \arrow[l,shift left=.8ex,color=red,"k_{-4}"]
&
X_9+X_{12}
  \arrow[r,shift left=.8ex,"k_5"]
&
X_{13}
  \arrow[l,shift left=.8ex,color=red,"k_{-5}"]
  \arrow[r,color=red,"\kappa_5"]
&
X_9+X_{14}
&
X_{14}+X_{15}
  \arrow[r,shift left=.8ex,"k_6"]
&
X_{16}
  \arrow[l,shift left=.8ex,color=red,"k_{-6}"]
  \arrow[r,color=red,"\kappa_6"]
&
X_{14}+X_{17}
\\
 X_{17}+X_7
 \arrow[r,shift left=.8ex,"k_7"]
 &
 X_{18} 
 \arrow[l,shift left=.8ex,color=red,"k_{-7}"]
  \arrow[r,color=red,"\kappa_7"]
 &
 X_{17}+X_9
 & 
 X_{14}+X_{19}
 \arrow[r,shift left=.8ex,"k_8"]
 &
 X_{20} 
 \arrow[l,shift left=.8ex,color=red,"k_{-8}"]
  \arrow[r,color=red,"\kappa_8"]
 &
 X_{58}+X_{19}
\\
 X_{14}+X_{21}
 \arrow[r,shift left=.8ex,"k_9"]
 &
 X_{22} 
 \arrow[l,shift left=.8ex,color=red,"k_{-9}"]
  \arrow[r,color=red,"\kappa_9"]
 &
 X_{14}+X_{23}
 &
 X_9+X_{24}
 \arrow[r,shift left=.8ex,"k_{10}"]
 &
 X_{25} 
 \arrow[l,shift left=.8ex,color=red,"k_{-10}"]
  \arrow[r,color=red,"\kappa_{10}"]
 &
 X_9+X_{26}
\\
 X_{26}+X_{27} 
 \arrow[r,shift left=.8ex,"k_{11}"]
 & 
 X_{28}
 \arrow[l,shift left=.8ex,color=red,"k_{-11}"]
 &
 X_{26}+X_{29}
 \arrow[r,shift left=.8ex,"k_{12}"]
 &
 X_{30} 
 \arrow[l,shift left=.8ex,color=red,"k_{-12}"]
  \arrow[r,color=red,"\kappa_{12}"]
 &
 X_{26}+X_{31}
 & 
 X_{31} 
 \arrow[r,shift left=.8ex,color=red,"k_{13}"]
 &
 X_{32}
 \arrow[l,shift left=.8ex,color=red,"k_{-13}"]
\\
 X_{32} + X_{33} 
 \arrow[r,shift left=.8ex,"k_{14}"]
 &
 X_{34}
 \arrow[l,shift left=.8ex,color=red,"k_{-14}"]
 &
 2 X_{32}  
 \arrow[r,shift left=.8ex,"k_{15}"]
 &
 X_{35}
 \arrow[l,shift left=.8ex,color=red,"k_{-15}"]
 &
 X_{35}+X_{33}  
 \arrow[r,shift left=.8ex,"k_{16}"]
 &
 X_{36}
 \arrow[l,shift left=.8ex,color=red,"k_{-16}"]
\\
 2 X_{35} 
 \arrow[r,shift left=.8ex,"k_{17}"]
 &
 X_{37}
 \arrow[l,shift left=.8ex,color=red,"k_{-17}"]
 & 
 X_{37} + X_{33} 
 \arrow[r,shift left=.8ex,"k_{18}"]
 &
 X_{38}
 \arrow[l,shift left=.8ex,color=red,"k_{-18}"]
 &
 X_{37}+X_{39}
 \arrow[r,shift left=.8ex,"k_{19}"]
 &
 X_{40} 
 \arrow[l,shift left=.8ex,color=red,"k_{-19}"]
  \arrow[r,color=red,"\kappa_{19}"]
 &
 X_{41}
\\
 X_{41}+X_{42}
 \arrow[r,shift left=.8ex,"k_{20}"]
 &
 X_{43} 
 \arrow[l,shift left=.8ex,color=red,"k_{-20}"]
  \arrow[r,color=red,"\kappa_{20}"]
 &
 X_{41}+X_{44}
 & 
 X_{41}+X_{45}
 \arrow[r,shift left=.8ex,"k_{21}"]
 &
 X_{46} 
 \arrow[l,shift left=.8ex,color=red,"k_{-21}"]
  \arrow[r,color=red,"\kappa_{21}"]
 &
 X_{41}+X_{47}
 & 
 X_{44} 
 \arrow[r,shift left=.8ex,color=red,"k_{22}"]
 &
 X_{48}
 \arrow[l,shift left=.8ex,color=red,"k_{-22}"]
\\
 X_{48}+X_{49}
 \arrow[r,shift left=.8ex,"k_{23}"]
 &
 X_{50} 
 \arrow[l,shift left=.8ex,color=red,"k_{-23}"]
  \arrow[r,color=red,"\kappa_{23}"]
 &
 X_{48}+X_{51}
 &
 X_{51}+X_{52} 
 \arrow[r,shift left=.8ex,"k_{24}"]
 &
 X_{53}
 \arrow[l,shift left=.8ex,color=red,"k_{-24}"]
 & 
 X_{53}+X_{12}
 \arrow[r,shift left=.8ex,"k_{25}"]
 &
 X_{54} 
 \arrow[l,shift left=.8ex,color=red,"k_{-25}"]
  \arrow[r,color=red,"\kappa_{25}"]
 & 
 X_{53}+X_{14}
\\
 X_{47} 
 \arrow[r,shift left=.8ex,color=red,"k_{26}"]
 &
 X_{55}
 \arrow[l,shift left=.8ex,color=red,"k_{-26}"]
 &
 X_{53}+X_{19} 
 \arrow[r,shift left=.8ex,"k_{27}"]
 &
 X_{56}
 \arrow[l,shift left=.8ex,color=red,"k_{-27}"]
 &
 X_{55}+X_{19} 
 \arrow[r,shift left=.8ex,"k_{28}"]
 &
 X_{57}
 \arrow[l,shift left=.8ex,color=red,"k_{-28}"]
\end{tikzcd}
\end{adjustbox}
\caption{The 
reaction network $\widetilde{N}$ underlying the Albeck {\em et al.}\ model~\cite{albeck2008}.
Removing all zero-order and first-order reactions, which are shown in red, yields the subnetwork $N$.
} \label{fig:EARM}
\end{figure}

\medskip

Consider now the subnetwork $N$ obtained by removing all the zero-order and first-order reactions from the network $\widetilde{N}$ in Figure~\ref{fig:EARM}.  It turns out that $N$ has $33$ (linearly independent) conservation laws, all of which arise from stoichiometry; the fact that the linear conservation laws of $N$ can be generated by those arising from stoichiometry follows from the observation that each linkage class of $N$ has only one terminal strong linkage class~\cite{feinberg-horn-1977}.
Of these $33$ conservation laws, $15$ of them are lifted to affine-linear steady-state relations of $\widetilde{N}$ via the map $\phi$ from Theorem~\ref{thm:main}.
These $15$ relations $h_1,h_2,\dots, h_{15}$ are shown in Table~\ref{tab:lifted}.} 
Moreover, equation~\eqref{eq:cor-dim-equation} in Corollary~\ref{coro:dimensionsum} holds, as follows: 
\[
  \underset{\dim_{\mathbb{Q}(\kappa)}(L_{\kappa}(N))}{\underbrace{33}} \, = \, \underset{\dim_{\mathbb{Q}({\kappa})} (A_{{\kappa}}(\widetilde{N}))}{\underbrace{15}}+\underset{\dim_{\mathbb{Q}(\kappa)}(L_\kappa(\widetilde{N}))}{\underbrace{18}}.
\]

We claim that Hypothesis~\ref{hyp:1} holds for the networks $N$ and $\widetilde{N}$.  
The second part of the hypothesis holds, as we already saw that each linkage class of $N$ has only one terminal strong linkage class (as in 
Remark~\ref{rem:hyp-1-part-2}).
Next, $\widetilde{N}$ does not contain the $0$ complex, so we consider one of the monomolecular reactant complexes $X_j$ of $\widetilde{N}$.  Observe that $X_j$ has at most two removed reactions (shown in red in Figure~\ref{fig:EARM}) coming out of it, and that if there are two such reactions (as in the case of $X_j = X_3$), one of the reactions is the reverse of a reaction of $N$ and hence its reaction vector belongs to $S_N$.  We conclude that Hypothesis~\ref{hyp:1} holds. 
Thus, we know by Proposition~\ref{prop:Hi-plus} that the $15$ relations $h_1,h_2,\dots, h_{15}$ shown in Table~\ref{tab:lifted} are linearly independent over $\mathbb{Q}(\widetilde{\kappa})$.

\begin{table}[htb!]
 {\small
 \begin{align*}
  h_1 &= x_{3}, & h_2 &= x_{16}, & h_3 &= x_{20}, & h_4 &= x_{22}, \\
  h_5 &= x_{25}, & h_6 &= x_{30}, & h_7 &= x_{40}, & h_8 &= x_{43}, \\
  h_9 &= x_{46}, & h_{10} &= x_{50}, & h_{11} &= \textcolor{magenta}{\frac{\kappa_{25}}{\kappa_5}x_{54} + x_{13}},& h_{12} &= \textcolor{magenta}{\frac{\kappa_3}{\kappa_7}x_8+x_{18}}, \\
  h_{13} &= x_{31}-\frac{k_{-13}}{k_{13}}x_{32}, & h_{14} &= x_{44}-\frac{k_{-22}}{k_{22}}x_{48}, &  h_{15} &= x_{47}-\frac{k_{-26}}{k_{26}}x_{55}. & & 
 \end{align*}
 }
\caption{The affine-linear steady state relations of $\widetilde{N}$, obtained by the map~$\phi$ from the subnetwork $N$ without zero-order and first-order reactions. Shown in magenta are the two relations that are sums of two positive terms. 
}\label{tab:lifted}

\end{table}

\medskip

Our next aim is to use the affine-linear steady-state relations to reduce the number of variables. For instance, if such a relation has the form $h_i=x_j$, then $x_j^*=0$ at any steady state $x^*$. More generally, we apply Lemma~\ref{lem:positive_rel} to the relations $h_1,h_2, \dots, h_{12}$ in Table~\ref{tab:lifted}, and conclude that $14$ variables vanish at every non-negative steady state: 
\begin{align}\label{eq:14-are-zero}
    x_3=\textcolor{magenta}{x_8}=\textcolor{magenta}{x_{13}}=x_{16}=\textcolor{magenta}{x_{18}}=x_{20}=x_{22}=x_{25}=x_{30}=x_{40}=x_{43}=x_{46}=x_{50}=\textcolor{magenta}{x_{54}}=0~. 
\end{align}    
    The four variables shown above in magenta ($x_8,x_{13},x_{18},x_{54}$) occur in the linear relations $h_{11}$ and $h_{12}$ that involve two terms with \emph{positive} coefficients (as in the form in Lemma~\ref{lem:positive_rel}).

Finally, the remaining three relations in Table~\ref{tab:lifted} (namely, $h_{13}$, $h_{14}$, and $h_{15}$) allow us to express, at steady state, $3$ variables as multiples of the others: 
\begin{align*}
x_{31}=\frac{k_{-13}}{k_{13}}x_{32}~, \quad  x_{44}=\frac{k_{-22}}{k_{22}}x_{48}~, \quad  x_{47}=\frac{k_{-26}}{k_{26}}x_{55}~.
\end{align*}

\medskip

To summarize, the $15$ affine-linear steady state relations allow us to reduce  the original system of steady-state equations in $58$ variables, to a system in $41$ variables. 

\subsection{Boundary steady states}  \label{sec:albeck-boundary}
With the aid of Gr\"obner-basis computations (see supplementary {\tt Maple} file), an additional $11$ variables can be expressed as a multiple of other variables, and the resulting steady-state equations admit a generating set consisting of monomials, as follows.

\begin{lemma}
[Steady states satisfy monomial equations]
    \label{lem:monomial-albeck}
Consider the Albeck~{\em et al.} network in Figure~\ref{fig:EARM}.  For any choice of positive rate constants $\kappa \in \mathbb{R}^{70}_{>0}$, 
every  non-negative steady state vanishes at the following $14$ monomials: 
\[
x_1x_2,
~ x_4x_7,
~ x_7 x_{17},
~ x_9 x_{12},
~ x_{12}x_{51}x_{52},
~ x_{14}x_{15},
~ x_{14}x_{19},
~ x_{14}x_{21},
~ x_9 x_{24},
~ x_{26}x_{29},
~ x_{32}^4x_{39},
~ x_{41}x_{42},\]
\[ x_{41}x_{45},
~x_{48}x_{49}~.
\]
Thus, all non-negative steady states lie on the coordinate hyperplanes (and so are boundary steady states).
\end{lemma}

The conservation laws involving the total amounts of caspase-6 (C6) and DNA damage repair enzyme (PARP) are $x_{15}+x_{16}+x_{17}+x_{18}=T_{C_6}$, and $x_{21}+x_{22}+x_{23}=T_{PARP}$ (earlier, $T_{C_6}$ and $T_{PARP}$ were called $T_7$ and $T_9$, respectively).
Recall from~\eqref{eq:14-are-zero} that $x_{16}=x_{18}=x_{22}=0$ holds at any non-negative steady state; thus, the following equations also hold at any non-negative steady state: 
\begin{align}
    \label{eq:2-equations-boundary}
x_{15}+x_{17}=T_{C_6} \quad {\rm and } \quad 
x_{21}+x_{23}=T_{PARP}~.
\end{align}
The variables $x_{15}, x_{17}, x_{21}$ and $x_{23}$ represent [C6], [C6$^*$], and the concentrations of PARP and cPARP (DNA damage repair enzyme and the cleaved form, which is a measure of cell death), respectively. 

After making all the above-mentioned substitutions into the steady-state equations and conservation-law equations, only two equations 
involve any of the following four variables: $x_{15}, x_{17}, x_{21}$, and $x_{23}$. These two equations are the ones shown in~\eqref{eq:2-equations-boundary}. 
Recall that we already saw that, for any choice of positive rate constants, every stoichiometric compatibility class contains at least one non-negative steady state.  
Thus, the intersection of the set of steady states with each stoichiometric compatibility class $\mathcal{P}$ 
that intersects the positive orthant~$\mathbb{R}_{>0}^{58}$ has dimension at least two. Here we use the fact that the total amounts 
$T_{C_6}$ and $T_{PARP}$ are strictly positive, which follows from the assumption $\mathcal{P} \cap \mathbb{R}_{>0}^{58} \neq \varnothing$. In fact, it can be shown that this intersection (of the set of steady states with some $\mathcal{P}$) has exactly dimension two for a range of positive rate constants. (See supplementary {\tt Maple} file.)

Our final aim is to show that every steady state is degenerate.\footnote{The fact that all steady states, regardless of the rate constants, are degenerate does not contradict the main result in~\cite{FHP25}, as their work concerns positive steady states, while all steady states of the Albeck {\em et al.}\ model here are non-positive (boundary steady states).}
In the Albeck {\em et al.} model, the variable $x_{14}$ represents [C3$^*$], the concentration of the active form of caspase-3. This active form acts as an enzyme with both C6 ($x_{15}$) and PARP $(x_{21})$. In turn, $x_{19}$ represents the concentration of XIAP, an inhibitor of apoptosis in the cell. This dual behavior of C3$^*$ is described in the following reactions:

$C3^* + C6 \overset{k_6}{\underset{k_{-6}}{\rightleftarrows}} C3^*:C6 \overset{\kappa_6}{\longrightarrow} C3^* + C6^*$, 
$C3^* + PARP \overset{k_9}{\underset{k_{-9}}{\rightleftarrows}} C3^*:PARP \overset{\kappa_9}{\longrightarrow} C3^* + cPARP$,
$C3^* + XIAP \overset{k_8}{\underset{k_{-8}}{\rightleftarrows}} C3^*:XIAP \overset{\kappa_8}{\longrightarrow} C3^*_{Ub}+ XIAP$.

\noindent
These interactions (plus other reactions in the network) yield the following linear relations, with non-constant coefficients, among the right-hand sides of the ODEs $\dot x_i= f_i^{\widetilde{N}}=:f_i $: 
\begin{align} \label{eq:syzygy-albeck}
\alpha_1 \, x_{15} \, f_{20} + \alpha_2 \, x_{15}  \, f_{58} + \alpha_3 \, x_{19} \, f_{16} + \alpha \, x_{19}  \, f_{17} + \alpha \, x_{19}  \, f_{18} ~=~ & 0\\
\notag
\beta_1 \, x_{15}  \, f_{22} + \beta_2 \,  x_{15}  \, f_{23} + \beta_3 \, x_{21}  \, f_{16} + \beta \, x_{21}  \, f_{17}  + \beta \, x_{21}  \, f_{18} ~=~ & 0~,
\end{align}
where the coefficients $\alpha_i,\alpha, \beta_i,\beta$ are as follows:  
$\alpha_1 = k_6\kappa_6\kappa_8(k_{-9}+\kappa_9)$,
$\alpha_2 = k_6\kappa_6(k_{-8}+\kappa_8)(k_{-9}+\kappa_9)$, 
$\alpha_3 =\kappa_6k_8\kappa_8(k_{-9}+\kappa_9)$, 
$ \alpha = -k_8\kappa_8(k_{-6}+\kappa_6)(k_{-9}+\kappa_9)$, 
$ \beta_1 = k_6\kappa_6\kappa_9(k_{-8}+\kappa_8)$,  $ \beta_2 = k_6\kappa_6(k_{-8}+\kappa_8)(k_{-9}+\kappa_9)$, $\beta_3 = -\kappa_6k_9\kappa_9(k_{-8}+\kappa_8)$, and $\beta = k_9\kappa_9(k_{-6}+\kappa_6)(k_{-8}+\kappa_8)$.

\medskip

We checked that $x_{19}$ is nonzero at every non-negative steady state (see supplementary {\tt Maple} file), so we can apply Proposition~\ref{prop:syzygy} to the first equation in~\eqref{eq:syzygy-albeck}. Thus, this equation, when  evaluated at steady state, yields a non-trivial linear relation among the gradients of the $f_i$.
Additionally, it is straightforward to check that this relation can {\em not} be deduced from the linear conservation laws.
Therefore, at every non-negative steady state, the rank of the Jacobian of the EARM model is strictly smaller than the expected rank, $40=58-18$, which is the number of linearly independent polynomials from among the $f_1,f_2,\dots, f_{58}$. 
We conclude that every non-negative steady state is degenerate.

\smallskip

We summarize the previous discussion in the following theorem.

\begin{theorem}[Steady states of Albeck {\em et al.}\ model]
    \label{thm:albeck}
Consider the Albeck~{\em et al.} network in Figure~\ref{fig:EARM}.  For any choice of positive rate constants $\kappa \in \mathbb{R}^{70}_{>0}$, the following hold:
\begin{enumerate}
    \item Every non-negative steady state is a degenerate, boundary steady state.
    \item Every stoichiometric compatibility class $\mathcal{P}$ intersecting the positive orthant is compact, and the set of (boundary) steady states in $\mathcal{P}$ is positive-dimensional, and additionally there exist rate constants $\kappa^* \in \mathbb{R}^{70}_{>0}$ for which this set is $2$-dimensional.
\end{enumerate}
\end{theorem}

\section{Proofs} \label{sec:proofs}
This section contains the proofs of our results in Section~\ref{sec:linear}, except for Corollary~\ref{coro:dimensionsum} and Lemma~\ref{lem:positive_rel}, which have straightforward proofs so we omit them.

\medskip

We begin with a straightforward lemma.

\begin{lemma}\label{lem:affine-linear-comes-from-ell}
Assume that $N$ is a reaction network obtained from a reaction network $\widetilde{N}$ by removing \uline{all} zero-order and/or first-order reactions.  We follow the notation in Setup~\ref{setup:star} and Theorem~\ref{thm:main}. In particular, we write $f_i^{\widetilde{N}} =
f_i^N + \ell_i$, where $\ell_i$ is an affine linear form for all $i=1,2, \dots, n$.

If $d_1,d_2, \dots, d_n \in \mathbb Q(\widetilde{\kappa})$ are such that $d_1 f_1^{\widetilde{N}} + d_2 f_2^{\widetilde{N}} + \dots + d_n f_n^{\widetilde{N}} =0$, then
the following equalities hold:
    \begin{align}\label{eq:lemma-lin-combo}
        d_{1} f^N_1 + d_{2} f^N_2 + \dots + d_{n} f^N_n~=~0~
        \quad {\rm and}
        \quad 
         d_{1} \ell_1 + d_{2} \ell_2 + \dots + d_{n} \ell_n~=~0~.
    \end{align}
    \end{lemma}
The proof of Lemma~\ref{lem:affine-linear-comes-from-ell} is evident because the polynomial $f_i^N$ and the linear form $\ell_i$ have {\em distinct monomials} (of degree at least $2$ in the first case, and of degree at most $1$ in the second case) for any index $i=1,2, \dots, n$.  We conclude that both sums in~\eqref{eq:lemma-lin-combo} need to be zero; and we use this basic argument in what follows.

\begin{proof}[Proof of Theorem~\ref{thm:main}]
    Part~\eqref{it:affine-linear} follows easily from the fact that $\widetilde{N}$ is obtained from $N$ by adding 
 \zerothOrderRxnsSpace
and/or first-order reactions (such reactions yield monomials of degree $0$ or $1$ in the mass-action ODEs).  

Next, for part~\eqref{it:phi}, assume that $c \in L_{\kappa}(N)$. 
This means, by definition, that $c \in \mathbb{Q}(\kappa)^n$ and $c_1 f^N_1+c_2f^N_2+\dots +c_n f^N_n =0$.  Thus, using the fact that $\ell_i=f_i^{\widetilde{N}}-f_i^N$, we obtain:
\begin{align}
    \notag
    c_1 f_1^{\widetilde{N}}+c_2f_2^{\widetilde{N}}+\dots +c_n f_n^{\widetilde{N}}
    ~=~&    
    \left( c_1 f^N_1+c_2f^N_2+\dots +c_n f^N_n \right) 
    \\
    & \label{eq:proof-decompose-f}
    \quad +~ \left( c_1 \ell_1+c_2\ell_2+\dots +c_n \ell_n \right) \\ \notag 
    ~=~ & 0+
        \left( c_1 \ell_1+c_2\ell_2+\dots +c_n \ell_n \right) ~.
\end{align}

It now follows that  that if $f^{\widetilde{N}}(\widetilde{\kappa}^*, x^*)=0$, then  $h:=\phi(c) =c_1 \ell_1+c_2\ell_2+\dots +c_n \ell_n $ also vanishes at $(\widetilde{\kappa}^*, x^*)$, i.e., $h$ is a steady-state relation of $\widetilde{N}$.  Also, $h$ is affine since each $\ell_i$ is affine.  Finally, we see from~\eqref{eq:proof-decompose-f} that $h$ is obtained {\em linearly} from the $f_i^{\widetilde{N}}$, with coefficients from $\mathbb{Q}(\kappa)$.  We conclude that $h$ is in the space
$ A_{{\kappa}}(\widetilde{N})$, as in Notation~\ref{notation:N-subnet-N-tilde}.  Finally, the fact that the map $\phi$ is $\mathbb{Q}(\kappa)$-linear is straightforward. 
This completes part~\eqref{it:phi}.

Part~\eqref{it:kernel} concerns the kernel of the linear map $\phi: L_{\kappa}(N) \to A_{{\kappa}}(\widetilde{N}) $. Let $c \in L_{\kappa}(N)$ (i.e. $c \in \mathbb{Q}(\kappa)^n$ and $c_1f^N_1+c_2f^N_2+ \dots + c_nf^N_n=0$).  As above, in~\eqref{eq:proof-decompose-f}, we have $ c_1 \ell_1+c_2\ell_2+\dots +c_n \ell_n =c_1 f_1^{\widetilde{N}}+c_2f_2^{\widetilde{N}}+\dots +c_n f_n^{\widetilde{N}}$.
Thus, $c_1 \ell_1+c_2\ell_2+\dots +c_n \ell_n =0$ (i.e., $c \in \ker (\phi)$) if and only if $c_1 f_1^{\widetilde{N}}+c_2f_2^{\widetilde{N}}+\dots +c_n f_n^{\widetilde{N}} = 0$.  
In particular, if $c \in \ker(\phi)$, then $c \in L_{\kappa}(\widetilde{N})$; in other words, the containment 
$\ker (\phi) \subseteq L_{\kappa}(\widetilde{N})$ holds.  Finally, the remaining containment, $L_{\kappa}(\widetilde{N}) \subseteq L_{\widetilde{\kappa}}(\widetilde{N})$,  follows readily from Definition~\ref{def:spaces-conservation-laws} and Notation~\ref{notation:N-subnet-N-tilde}.

We now turn to part~\eqref{it:all}.  Accordingly, assume that $N$ is obtained from $\widetilde{N}$ by removing all zero-order and first-order reactions.
We begin with item~\ref{it:kernel-if-remove-all}.  
Based on the description of the kernel, in part~\eqref{it:kernel} of Theorem~\ref{thm:main}, it suffices to prove the following equality:
\begin{align*}
\{c \in L_{\kappa}(N) \mid c_1f_1^{\widetilde{N}} + \dots + c_nf_n^{\widetilde{N}} =~0\}
~=~
\{c \in \mathbb{Q}(\kappa)^n \mid c_1f_1^{\widetilde{N}} + \dots + c_nf_n^{\widetilde{N}} =~0\}~.
\end{align*}
The containment ``$\subseteq$'' follows from the fact that, by definition, $L_{\kappa}(N) \subseteq \mathbb{Q}(\kappa)^n$.  Next, for the containment ``$\supseteq$'', assume that $c \in \mathbb{Q}(\kappa)^n$ satisfies $c_1f_1^{\widetilde{N}}+ c_2f_2^{\widetilde{N}} + \dots + c_nf_n^{\widetilde{N}} =0$.

Thus, using $\ell_i=f_i^{\widetilde{N}}-f_i^N$, we obtain  $c_1 \ell_1 + c_2 \ell_2 + \dots + c_n \ell_n= -(c_1f_1^{N} + c_2f_2^{N} + \dots + c_nf_n^{N})$. The polynomials on each side of this equality have distinct monomials because the left-hand side has terms of degree $0$ and $1$, while the right-hand side has terms of degree $2$ or more because $N$ has no zero-order and first-order reactions. Thus, as in Lemma~\ref{lem:affine-linear-comes-from-ell}, both polynomials must be zero. 
In particular, $c_1 f_1^N  + \dots + c_n f_n^N=0$ and so $c \in L_\kappa(N)$.
We conclude that item~\ref{it:kernel-if-remove-all} holds.

We now prove item~\ref{it:surjective}.  
Let $h \in  A_{{\kappa}}(\widetilde{N})$.  
This implies that $h = c_1 f_1^{\widetilde{N}} + c_2 f_2^{\widetilde{N}} + \dots + c_n f_n^{\widetilde{N}}$ for some $c_i \in \mathbb{Q}({\kappa})$.
Also, as $A_{{\kappa}}(\widetilde{N}) \subseteq A_{\widetilde{\kappa}}(\widetilde{N})$, we have that $h \in A_{\widetilde{\kappa}}(\widetilde{N})$.  
Then by the argument behind Lemma~\ref{lem:affine-linear-comes-from-ell}, $c_1 f^N_1 + c_2 f^N_2 + \dots + c_n f^N_n=0$, that is, $c \in L_{\kappa}(N)$,  and so $\phi(c)$ is defined and equals $ c_1 \ell_1 + c_2 \ell_2 + \dots + c_n \ell_n = h$.  Thus, as desired, $h$ is in the image of $\phi$. So,  $\phi$ is surjective and item~\ref{it:surjective} holds. 

\end{proof}

\medskip

\begin{proof}[Proof of Proposition~\ref{prop:generates}]
Assume that $N$ is obtained from $\widetilde{N}$ by removing all zero-order and first-order reactions. 
We must show that $A_{\widetilde{\kappa}}(\widetilde{N})=\mathrm{span}_{\mathbb{Q}(\widetilde{\kappa})}(A_{\kappa}(\widetilde{N}))$. The containment ``$\supseteq$'' is straightforward from definitions. For the reverse containment, consider some $h \in A_{\widetilde{\kappa}}(\widetilde{N})$.   
By definition, this means the following:
    \begin{itemize}
        \item $h = d_1 f_1^{\widetilde{N}} + d_2 f_2^{\widetilde{N}} + \dots + d_n f_n^{\widetilde{N}}$ for some $d_i \in \mathbb{Q}(\widetilde{\kappa})$, and
        \item when simplified, $h$ has only terms of degree $0$ and $1$ in $x$.
    \end{itemize}
The argument behind Lemma~\ref{lem:affine-linear-comes-from-ell} now implies the following: 
\begin{align} \label{eq:proof-lin-combo-N2}
    d_{1} f^N_1 + d_{2} f^N_2 + \dots + d_{n} f^N_n~=~0~
        \quad {\rm and}
        \quad 
        h ~=~ d_{1} \ell_1 + d_{2} \ell_2 + \dots + d_{n} \ell_n~.
    \end{align}
Next, the entries of $d:=(d_1,d_2,\dots, d_n)^{\top} \in \mathbb{Q}(\widetilde{\kappa})^n$ may contain denominators.  By scaling $d$ by a common denominator (e.g., the product of the denominators of the $d_i$), we may assume that $d \in \mathbb{Q}[\widetilde{\kappa}]^n$.  
We also scale $h$ by the same common denominator, and so the equalities~\eqref{eq:proof-lin-combo-N2} remain true (and observe that this rescaled $h$ still has only degree-0 and degree-1 terms in $x$).  Finally, as we are viewing $A_{\kappa}(\widetilde{N})$ as a $\mathbb{Q}(\widetilde{\kappa})$-vector space, the scaling we did to $h$ does not affect whether or not $h$ is in $\mathrm{span}_{\mathbb{Q}(\widetilde{\kappa})}(A_{\kappa}(\widetilde{N}))$.  So, we can consider the rescaled $h$ (which, for simplicity, we continue to call $h$) in the rest of the proof.
    
We now rewrite the vector $d$ in terms of the $p$ ``new'' rate constants $\kappa':= (k_{r+1}, k_{r+2}, \dots, k_{r+p})$, as follows:
    \begin{align}
        \label{eq:write-d-in-terms-new-kappa}
        d ~=~ \sum_{a \in A} (\kappa')^a D(a)~,
    \end{align}
where $A$ is a (fixed) finite subset of $\mathbb{Z}^p_{\geq 0}$ and, for all $a \in A$, the vector $D(a)$ involves only ``old'' rate constants (more precisely, $D(a)  \in \mathbb{Q}[\kappa]^n$).  
    
For ease of notation, we use $f^N:=(f^N_1, f^N_2, \dots, f^N_n)^{\top}$ below.  Using~\eqref{eq:proof-lin-combo-N2}--\eqref{eq:write-d-in-terms-new-kappa}, we obtain:
    \begin{align}
        \label{eq:use-rewritten-d}
        0   ~=~ \langle d, ~f^N \rangle 
            ~=~ \sum_{a \in A} ~(\kappa')^a~ \langle D(a),~f^N \rangle~.
    \end{align}
In the above expression, each term $ \langle D(a),~ f^N \rangle$ is a polynomial in the ``old'' rate constants (that is, $ \langle D(a)~, f^N \rangle \in \mathbb{Q}[\kappa][x]$) since the entries of both vectors $D(a)$ and $f^N$ are such polynomials.  Therefore, in order for the sum on the right-hand side of~\eqref{eq:use-rewritten-d} to be the zero polynomial, it must be that 
the following equality holds for all $a \in A$:
\begin{align}
    \label{eq:proof-D(a)}
    \langle D(a), f^N \rangle~=~0~.
\end{align} 

Call $\ell:=(\ell_1, \ell_2, \dots, \ell_n)^{\top}$, then by~\eqref{eq:proof-lin-combo-N2}--\eqref{eq:write-d-in-terms-new-kappa}, we have:
\[
 h~=~\langle d,\ell\rangle ~=~ \sum_{a \in A} ~(\kappa')^a~ \langle D(a),~\ell \rangle~.
\]
Recall that our goal is to show that $h$ is in the $\mathbb{Q}(\widetilde{\kappa})$-span of $A_{\kappa}(\widetilde{N})$.  Therefore, in light of the above equation, it suffices to show that, for all $a \in A$, the expression
$h_a:=\langle D(a),~\ell \rangle$ is in $A_{\kappa}(\widetilde{N})$, as in Notation~\ref{notation:N-subnet-N-tilde}.  
We confirm this as follows.  First, $h_a$ is affine linear, because $D(a) \in \mathbb{Q}[\kappa]^n$ and each $\ell_i$ is affine linear.  Next, we  
use~\eqref{eq:proof-D(a)} to obtain the following:
\[
h_a ~=~ \langle D(a),~\ell \rangle 
~=~ \langle D(a),~\ell \rangle + \langle D(a),~f^N \rangle 
~=~  \langle D(a),~f^{\widetilde{N}} \rangle ~.
\]
It now readily follows that $h\in \mathrm{span}_{\mathbb{Q}(\widetilde{\kappa})}(A_{\kappa}(\widetilde{N}))$, as desired.
\end{proof}

\smallskip

\begin{proof}[Proof of Proposition~\ref{prop:syzygy}]
Fix $\kappa^*$ as in the statement of the proposition.  Assume that $x^*$ is a common zero of the polynomials $f_i$ (here and in what follows, we omit the reference to $\kappa^*$ to alleviate the notation).
Differentiating the identity $\sum _{i=1}^m g_i \, f_i =0 $ gives $\sum _{i=1}^m \frac {\partial }{\partial x_j}(g_i \, f_i) =0 $ for any $j=1,2, \dots, n$.
Next, for all $i=1,2,\dots,m$ and all $j =1,2 \dots, n$:
\[ 
    \left( \frac{\partial}{\partial x_j} (g_i f_i) \right) (x^*)~=~
\frac{\partial g_i(x^*)}{\partial x_j} f_i(x^*)+
g_i (x^*) \frac{\partial f_i(x^*)} {\partial x_j} ~=~
0+ g_i (x^*) \frac{\partial f_i(x^*)} {\partial x_j}~.
\]
The desired linear dependence~\eqref{eq:dependence-gradient} follows and is nontrivial when
 not all $g_i$ vanish at $x^*$.
\end{proof}

\medskip

Before we prove Proposition~\ref{prop:Hi-plus}, we fix some notation to be able to describe the coefficients of any $h\in A_\kappa(\widetilde{N})$.

\begin{notation} \label{notation:proofs-involving-hypotheses1-2}
We set $x_0:=1$. For $h\in A_{\kappa}(\widetilde{N})$, we write $h=a_0x_0+a_1x_1+a_2x_2+\dots+a_nx_n$ where $a_j\in\mathbb{Q}(\widetilde{\kappa})$ for $j=0,1,2,\dots,n$.  We let $\mathcal{I}_j$ denote the set of indices of the reactions in which the reactant complex is the complex $X_j$ (and we write $\mathcal{I}_0$ if the reactant complex is the zero complex). 
Finally, we denote the reaction vector of the reaction $y_i \to y_i'$ by $v_i:=y_i'-y_i$.
\end{notation}

\begin{proof}[Proof of Proposition~\ref{prop:Hi-plus}]

By Proposition~\ref{prop:generates}, we know $A_{\widetilde{\kappa}}(\widetilde{N})=\mathrm{span}_{\mathbb{Q}(\widetilde{\kappa})}(A_{\kappa}(\widetilde{N}))$, and as $\mathbb{Q}(\widetilde{\kappa})$ is an extension field of $\mathbb{Q}(\kappa)$, we have
\begin{equation}\label{eq:less_dim}
    \dim_{\mathbb{Q}(\widetilde{\kappa})}(A_{\widetilde{\kappa}}(\widetilde{N}))~\leq~ \dim_{\mathbb{Q}(\kappa)}(A_{\kappa}(\widetilde{N}))~.
\end{equation}
We will show that under Hypotheses~\ref{hyp:1} or~\ref{hyp:2}, the inequality in~\eqref{eq:less_dim} is an equality. 
 
\medskip
 
Assume Hypothesis~\ref{hyp:1} is satisfied by $N$ and $\widetilde{N}$. Let $s:=\dim_{\mathbb{Q}(\kappa)}(A_{\kappa}(\widetilde{N}))$, and consider a $\mathbb{Q}(\kappa)$-basis of affine-linear functions $h^1,h^2,\dots,h^s$ of $A_{\kappa}(\widetilde{N})$. By Theorem~\ref{thm:main}\eqref{it:all}(ii), $\phi$ is surjective, so,  for each $1\leq l\leq s$, there exists $d^l \in L_\kappa(N)$ such that  $h^l= \phi(d^l)=\sum_{i=1}^n d_i^l f_i^{\widetilde{N}}$.
Using Notation~\ref{notation:proofs-involving-hypotheses1-2}  and~\eqref{eq:mass_action_ODE}, we have $f_j^{\widetilde{N}}=\sum_{i=1}^{r+p}  k_{i} x^{y_i} v_{ij}$ and $h^l =\sum_{i=1}^{r+p}  k_{i} x^{y_i} \langle d^l,~ v_i \rangle = \sum_{j=0}^n x_j \sum_{i\in \mathcal{I}_j}k_i \langle d^l,v_i\rangle$. This way, we obtain the following:
\begin{equation}\label{eq:h_by_reactions}
  h^l=\sum_{j=0}^n a^l_jx_j~,
  \quad 
  \text{ with } a^l_j=\sum_{i\in \mathcal{I}_j}k_i \langle d^l,v_i\rangle~,
\end{equation}
for $1\leq l\leq s$. 
The fact that the $h^1,h_2, \dots, h^s$ in $A_{\widetilde{\kappa}}(\widetilde{N})$ 
are linearly independent can be readily translated into a condition on the rows
of the matrix of coefficients $(a_j^l)$, where the superscript $l$ indexes rows and the subscript $j$ indexes columns.  

We first consider the case in which, for each $0\leq j\leq n$, there exists one reaction out of $X_j$, indexed by some $i_j\in\mathcal{I}_j$ such that $v_{i_j}\notin S_N$. In this case, $\langle d^l,v_i\rangle=0$ for all $i\in \mathcal{I}_j\setminus \{i_j\}$, where $1\leq l\leq s$. The matrix of coefficients of $h^1,h^2,\dots,h^s$ can then be factored as the product of the matrix $(\langle d^l,v_{i_j}\rangle) \in \mathbb{Q}(\kappa)^{s\times (n+1)}$, where again the superscript $l$ indexes rows and the subscript $j$ indexes columns, and an invertible diagonal matrix with $k_{i_0},k_{i_1},\dots, k_{i_n}$ in its diagonal, as follows:
\[
  \begin{pmatrix}
   \langle d^1,v_{i_0}\rangle & \langle d^1,v_{i_1}\rangle & \dots & \langle d^1,v_{i_n}\rangle \\
   \langle d^2,v_{i_0}\rangle & \langle d^2,v_{i_1}\rangle & \dots & \langle d^2,v_{i_n}\rangle \\
   \vdots & & &\vdots  \\
   \langle d^s,v_{i_0}\rangle & \langle d^s,v_{i_1}\rangle & \dots & \langle d^s,v_{i_n}\rangle \\
  \end{pmatrix}
  \begin{pmatrix}
   k_{i_0} & & & \\
   & k_{i_1} & & \\
    & & \ddots & \\
    & & & k_{i_n}
  \end{pmatrix}.
\]
As $h^1,h^2,\dots,h^s$ are linearly independent over $\mathbb{Q}(\kappa)$, the matrix on the left, $(\langle d^l,v_{i_j}\rangle)$, has $s$ linearly independent rows over $\mathbb{Q}(\kappa)$.
Moreover, as the matrix on the left has entries in the smaller field $\mathbb{Q}(\kappa)$, its rows are also linearly independent over the larger field $\mathbb{Q}(\widetilde{\kappa})$. We conclude that equality~\eqref{eq:kk} holds. Finally, in the remaining case, when,  for some $j$, there is no $i_j$ with $v_{i_j}\notin S_N$, we just need to consider a submatrix.

\medskip

Assume now that Hypothesis~\ref{hyp:2} holds. Consider a basis, $d^1,d^2,\dots, d^t\in L_\kappa(N)$ and species $X_{j_1},X_{j_2},\dots,X_{j_t}$ such that, for each $1\leq l\leq t$ the only removed reaction out of $X_{j_l}$ is  the degradation reaction $X_{j_{l}} \to 0$, and $d^{l}_{j_{\ell}}\neq0$ and $d^{l}_{j_{l+1}}=d^{l}_{j_{l+2}}=\dots =d^{l}_{j_{s}}=0$. Call $h^l:= \phi(d^l) =\sum_{i=1}^n d_i^l f_i^{\widetilde{N}}$.
Without loss of generality, we can reorder the reactions and assume $X_{j_l}\overset{k_{j_l}}{\longrightarrow}0$. Then, $\mathcal{I}_{j_l}=\{j_l\}$, the corresponding reaction vector is $v_{j_l}=-e_{j_l}$, and $\langle d^l, v_{j_l}\rangle=-d^l_{j_l}$. This way, the coefficient of $x_{j_l}$ in $h^l$ in~\eqref{eq:h_by_reactions} is $-k_{j_l}d^l_{j_l}\neq 0$, and $\langle d^l, v_{j_m}\rangle=0$ for all $m \geq l+1$. In this case, the $t\times t$ submatrix of the matrix of coefficients whose columns correspond to $x_{j_1}, x_{j_2},\dots,x_{j_t}$ is a lower-triangular matrix which can be now factored as a lower-triangular matrix with $-d^1_{j_1}$, $-d^2_{j_2}$,\dots,$-d^t_{j_t}$ in the diagonal (thus, it has $t$ linearly independent rows over $\mathbb{Q}(\kappa)$) and an invertible diagonal matrix with $k_{j_1},k_{j_2},\dots, k_{j_t}$ in its diagonal\footnote{ An instance of this decomposition was shown earlier, at the end of Example~\ref{ex:eissing-follow-up-on-intro}.}:
\[
  \begin{pmatrix}
   -d^1_{j_1} & 0 & \dots & 0 \\
    * & -d^2_{j_2} & \ddots  & \vdots\\
    \vdots & \ddots & \ddots & 0\\
    * & \dots & * & -d^t_{j_t} \\
  \end{pmatrix}
  \begin{pmatrix}
   k_{j_1} & & & \\
   & k_{j_2} & & \\
    & & \ddots & \\
    & & & k_{j_t}
  \end{pmatrix}.
 \]
Thus, $h^1,h^2,\dots,h^t$ are also linearly independent over $\mathbb{Q}(\widetilde{\kappa})$.
Hence, 
\[
\dim_{\mathbb Q(\widetilde{\kappa})}A_{\widetilde{\kappa}}(\widetilde{N})\ \ge \ \dim_{\mathbb Q(\kappa)}L_\kappa(N)\ \ge \ \dim_{\mathbb Q({\kappa})}A_{{\kappa}}(\widetilde{N})~,
\]
where the last inequality holds because $A_{{\kappa}}(\widetilde{N})=\mathrm{im}(\phi)$ (by Theorem~\ref{thm:main}). Together with~\eqref{eq:less_dim}, this gives the equality of the dimensions in~\eqref{eq:kk}.
\end{proof}

\section{Discussion} \label{sec:discussion}

Steady states drive the long-term  dynamics of chemical reaction networks (and, more generally, the dynamics of autonomous systems of ODEs of the form $\dot{x}=f(x)$). In addition, unstable steady states are related to the occurrence of oscillations. In this work, we showed how 
the  conservation laws for smaller subnetworks of a given chemical reaction network can be used to reduce the number of variables and to describe the steady states of the (big) network. We showcased the utility of our results through two biologically relevant models of programmed cell death  that exhibit interesting behavior.

\subsubsection*{Acknowledgments} AD and MPM were
 partially supported by UBACYT 20020220200166BA and CONICET PIP 11220200100182CO, Argentina.
AS was partially supported by the National Science Foundation (DMS-1752672 and DMS-2533213) 
and the Simons Foundation (through a Simons Fellowship in Mathematics). AS thanks Megan Dowdell and Patricia Klein for helpful conversations.  

\subsubsection*{Data availability} 
This work has no associated data.

\subsubsection*{Code availability} 
Computations underlying some of our examples are shown in two supplementary {\tt Maple} files. 

\subsubsection*{AI disclosure} 
The authors used Claude to suggest comments to add to the supplementary {\tt Maple} files, and to check proofs and detect typos in a prior version of this work.

\bibliographystyle{plain}

\end{document}